\documentclass[prd,aps,
nofootinbib,
floatfix,
superscriptaddress]{revtex4}
\usepackage{graphicx}
\usepackage{epsfig}
\usepackage{rotating}
\usepackage{amssymb}
\usepackage{subfigure}
\usepackage{dsfont}
\usepackage{psfrag}
\usepackage{amsmath,euscript,array,mathrsfs}
\usepackage{axodraw}

\newcommand{\beq}{\begin{equation}}
\newcommand{\eeq}{\end{equation}}
\newcommand{\beqs}{\begin{eqnarray}}
\newcommand{\eeqs}{\end{eqnarray}}
\newcommand{\lsim}{\mathrel{\raisebox{-
.6ex}{$\stackrel{\textstyle<}{\sim}$}}}
\newcommand{\gsim}{\mathrel{\raisebox{-
.6ex}{$\stackrel{\textstyle>}{\sim}$}}}

\def\hbar{\hspace{0pt}\raisebox{1pt}{$-$} \hspace{-7pt} h}

\def\di{\mbox{d}}
\def\r{\rho}

\newcommand{\be}{\begin{equation}}
\newcommand{\ee}{\end{equation}}
\newcommand{\bea}{\begin{eqnarray}}
\newcommand{\eea}{\end{eqnarray}}

\def\lbldef#1#2{\expandafter\gdef\csname #1\endcsname {#2}}

\def\href#1#2{#2}

\newcommand{\ber}{\begin{eqnarray}}
\newcommand{\eer}{\end{eqnarray}}

\newcommand{\beqar}{\begin{eqnarray}}

\newcommand{\eeqar}{\end{eqnarray}}

\newcommand{\dsl}
  {\kern.06em\hbox{\raise.15ex\hbox{$/$}\kern-.56em\hbox{$\partial$}}}

\newcommand{\eeqarr}{\end{eqnarray}}
\newcommand{\ZZ}{{\rm \kern 0.275em Z \kern -0.92em Z}\;}

\def\CC{{\mathchoice
{\rm C\mkern-8mu\vrule height1.45ex depth-.05ex
width.05em\mkern9mu\kern-.05em}
{\rm C\mkern-8mu\vrule height1.45ex depth-.05ex
width.05em\mkern9mu\kern-.05em}
{\rm C\mkern-8mu\vrule height1ex depth-.07ex
width.035em\mkern9mu\kern-.035em}
{\rm C\mkern-8mu\vrule height.65ex depth-.1ex
width.025em\mkern8mu\kern-.025em}}}

\def\RR{{\rm I\kern-1.6pt {\rm R}}}

\def\ZZ{{\rm Z}\kern-3.8pt {\rm Z} \kern2pt}
\def\IB{\relax{\rm I\kern-.18em B}}
\def\ID{\relax{\rm I\kern-.18em D}}
\def\II{\relax{\rm I\kern-.18em I}}
\def\IP{\relax{\rm I\kern-.18em P}}

\newcommand{\bear}{\begin{eqnarray}}
\newcommand{\eear}{\end{eqnarray}}

\def\arctanh{~{\rm arctanh}~}

\def\n{\nu}
  
\def\r{\rho}                                     

\def\6{\partial}

\def\bea{\begin{eqnarray}}
\def\eea{\end{eqnarray}}

\def\beqx{\begin{displaymath}}
\def\eeqx{\end{displaymath}}

\newcommand{\bmat}{\left(\begin{array}}
\newcommand{\emat}{\end{array}\right)}

\def\n{\nu}

\def\r{\rho}

\def\bo{{\raise-.3ex\hbox{\large$\Box$}}}               
\def\face{{\raise.2ex\hbox{$\displaystyle \bigodot$}\mskip-2.2mu \llap {$\ddot
        \smile$}}}                                   
\def\>{\rangle}                                      
\def\<{\langle}                                      

\def\leftrightarrowfill{$\mathsurround=0pt \mathord\leftarrow \mkern-6mu
        \cleaders\hbox{$\mkern-2mu \mathord- \mkern-2mu$}\hfill
        \mkern-6mu \mathord\rightarrow$}        
\def\dvec#1{\vbox{\ialign{##\crcr
        \leftrightarrowfill\crcr\noalign{\kern-1pt\nointerlineskip}
        $\hfil\displaystyle{#1}\hfil$\crcr}}}           

\def\-{\hphantom{-}}

\begin{document}
\title{The decay constant of the holographic techni-dilaton and the 125 GeV boson}

\author{Daniel Elander}
\affiliation{Department of Theoretical Physics, Tata Institute of Fundamental Research, \\
Homi Bhabha Road, Mumbai 400 005, India}
\author{ Maurizio Piai}
\affiliation{Department of Physics, College of Science, Swansea University,
Singleton Park, Swansea, Wales, UK.}

\date{\today}


\begin{abstract}
We critically discuss the possibility that the 125 GeV boson recently discovered at the LHC
is the holographic techni-dilaton, a composite state emerging from a strongly-coupled 
model of electroweak symmetry breaking.
This composite state differs from the SM for three main reasons. Its decay constant is in general larger than the electroweak scale,
hence suppressing all the couplings to standard model particles with respect to an elementary Higgs boson,
with the exception of the coupling to photons and gluons, which is expected to be larger than the standard-model equivalent.

We discuss three classes of questions. Is it possible to lower the decay constant, by changing the geometry of the holographic model?
Is it possible to lower the overall scale of the strong dynamics,
by modifying the way in which electroweak symmetry breaking is implemented in the holographic model?
Is there a clear indication in the data that production mechanisms other than gluon-gluon fusion
have been observed, disfavoring models in which the holographic techni-dilaton has a  large decay constant?

We show that all of these questions are still open, given the present status of theoretical as well as
phenomenological studies, and that at present the techni-dilaton hypothesis yields a fit to the data which is either
as good as the elementary Higgs hypothesis, or marginally better, depending on what sets of data are used
in the fit. We identify clear strategies for future work aimed at addressing these three classes of open questions.

In the process, we also compute the complete scalar spectrum of the two-scalar truncation describing the GPPZ model, as well as the decay constant of the holographic techni-dilaton in this model.

\end{abstract}

\maketitle

\tableofcontents

\section{Introduction}

July 4th 2012 will be remembered as one of the most important moments
in the history of particle physics: the LHC experiments ATLAS and CMS 
announced the discovery of a new boson with mass $m_s\simeq 125$ GeV,
and decay and production rates compatible with the Higgs particle of the 
minimal version of the Standard Model~\cite{ATLASJuly2012,CMSJuly2012}.
The next step is now to understand what is the nature of this particle,
and in particular what are its interactions and its dynamical origin.
A number of studies appeared fitting the data and comparing to a number of possible
interpretations of the results, in terms of the elementary Higgs particle of the Standard Model, or
of alternative scenarios~\cite{LLS,Corbett:2012dm,GKRS,EY,Montull:2012ik,EGMT,CFKVZ,Bertolini:2012gu,MY}. 
All of the studies agree on the fact that the global fit is in substantial agreement with the
Standard Model, given present uncertainties. But  there is some tension in the data, in particular
due to the fact that all the experiments see a substantial enhancement of the number of 
two photon events, which means that the best fit to the data would prefer 
a modification of some of the couplings of the putative Higgs particle, in particular the coupling to photons.

This is  the very beginning of a process of precision measurements: the number of  new particles
produced and  analyzed is  small,
and the error bars in the various measurements are large. There is unambiguous 
evidence that the new particle decays into two photons, and into two (real or virtual) $Z$ or $W$ bosons.
All other possible decays have not been established firmly yet.
Hence it is still premature to draw conclusions about the nature of the new boson.
Yet, it is interesting to ask whether specific models, alternative to the minimal version of the SM,
are favored or disfavored by current trends in the data, because in doing so we may identify what is the best strategy to actually
test such alternatives, even  from a purely theoretical viewpoint.

A particularly interesting hypothesis about electroweak symmetry breaking is that it might 
arise from a new, strongly-coupled interaction, a scenario that is usually referred to as Technicolor (TC)~\cite{TC,reviews}.
The most naive realizations of this idea suffer from many phenomenological drawbacks, for example with 
precision electroweak physics~\cite{Peskin,Barbieri}, besides
the technical  difficulties involved with dealing with a strongly-coupled field theory.
It is widely accepted that a generic strongly-coupled model is already ruled out by indirect 
tests of the electroweak theory, well ahead of the direct searches at  LHC. 
An exception is represented by models in which the strongly-coupled dynamics is 
qualitatively very different from QCD, and approximately scale-invariant at scales just above the electroweak one.
This class of models  is generally referred to as walking technicolor (WTC)~\cite{WTC},
and it is known that the phenomenological arguments against generic TC models fail to disprove WTC models,
which hence provide a viable and appealing framework for new physics.

There are two more reasons why WTC models are special.
First of all, because the approximate scale invariance is spontaneously broken by the condensates appearing at
the electroweak scale, it is plausible that  a parametrically  light composite scalar state be present in  the low-energy  spectrum.
This is often referred to as a (techni-)dilaton, and has been the focus of a large number of studies~\cite{dilaton1,dilaton2,
dilaton3,dilatonpheno,dilaton4,dilaton5D,dilatonnew,EPdilaton,LP,ENP}.
In particular, it is known that if such a state exists its couplings are qualitatively very similar to those of the SM Higgs particle.
The second reason why this possibility is interesting is that 
the strongly-coupled, quasi-conformal, multi-scale dynamics of WTC is suitable to be treated on the basis of
the ideas of gauge-gravity dualities~\cite{AdSCFT,reviewAdSCFT}, 
and indeed a large number of studies in this direction  has appeared in the literature~\cite{stringS,stringWTC,stringWTC2,AdSTC}.

In this paper we focus on the hypothesis that the new particle discovered at the LHC is the dilaton
predicted by models inspired by  gauge-gravity dualities (holography), which we refer to as the holographic techni-dilaton,
and we critically discuss its properties, both from the theoretical point of view and in view of the experimental data.
The aim of the paper is to highlight three classes of open questions and future directions for research,
all of which may lead to unambiguously distinguish between the (composite holographic techni-)dilaton
and (elementary standard-model) Higgs particle.

Arguments based on naive dimensional analysis (NDA) indicate that the presence of a light  dilaton in the spectrum of a strongly-coupled theory is possible, but is not a generic feature~\cite{Rattazzi-Zaffaroni}. The presence of a light techni-dilaton emerges only in special subclasses of strongly-coupled models, in which the strong dynamics itself produces a mass for the dilaton that is lower than its NDA estimates. In short, this is ensured only in models in which the formation of non-trivial condensates in the IR plays a crucial role in breaking scale invariance and triggering confinement. Finding models with such dynamical features is highly non-trivial in strongly-coupled field theories. This is the main motivation to use gauge-gravity dualities, and requires studying models that go beyond the most simple realizations of five-dimensional electroweak symmetry breaking based on effective field-theory arguments, such as the Randall-Sundrum model and its variations. In particular, in order to compute the mass of the techni-dilaton one must focus on models in which the full dual background can be derived from a string-theory construction. A noticeable example of such a case has been discussed in~\cite{ENP}, where a class of models obtained by compatifying Type IIB supergravity on $T^{1,1}$ has been studied and shown to admit a light scalar in the spectrum, the mass of which is controlled by the integration constants in the background. In this paper, we discuss other classes of models, also inspired by string-theory, but not as complicated (and as a consequence not as complete) as the one in~\cite{ENP}. We use these as toy models, which allow us to discuss other important phenomenological properties, besides the mass of the dilaton itself. We will identify regions of parameter space where a light dilaton exists, without further questioning this result, on the basis of the fact that it persists also in some more complete string-theory models~\cite{ENP}, and of the fact that the experimental data suggests that it is only this class of strongly-coupled models, and not generic ones, that is of interest. We then study the phenomenology of such models  in the relevant regions of parameter-space.

The paper is organized as follows.
We start by recalling briefly a few useful elements about the treatment of scalar composite states in 
the context of five-dimensional sigma-models coupled to gravity.
We then devote three sections to the three main classes of open questions we want to highlight.

First, we discuss the relation between the holographic techni-dilaton decay constant $F$,
the overall mass scale of the theory $\Lambda_0$ and the five-dimensional geometry.
As we will see with several examples, it turns out that  the numerical result for the decay constant is 
remarkably universal, yielding $F\simeq \sqrt{3/2} \Lambda_0$.
As a special example of a model relevant to these scenarios, 
we also complete the calculation of the spectrum of scalar excitations 
of the GPPZ model~\cite{GPPZ}, focusing on the light states in the whole parameter space,
 hence complementing the literature on the subject~\cite{GPPZspectrum}.
We leave a very non-trivial open question: what would happen if the background geometry deviated significantly from AdS over a large region, for instance exhibiting hyperscaling violation~\cite{hyperscaling}? Notice that a complete string-theory model, in which a parametrically light scalar such as the dilaton is present, has been identified~\cite{ENP}. However, this model has a five-dimensional metric that is not
AdS, but rather exhibits hyperscaling violation. It would be very interesting to know 
what the decay constant is in this and similar cases. This highly non-trivial possibility has not been explored yet in the literature.

The second question we want to pose is related to the relation between 
the fundamental scale of the theory and the bounds from precision electroweak physics, in particular related to the $S$ parameter of~\cite{Peskin}.
Aside from a few exceptions (for example the  phenomenological study in~\cite{FPV}), 
in the vast majority of the studies of models 
inspired by holography, electroweak symmetry breaking is induced by effects that take place deep in the IR
of the geometry. In the bottom-up approach~\cite{AdSTC, EPdilaton, LP} this is usually done either by 
imposing non-trivial boundary conditions in the deep IR, or by assuming that a bulk Higgs field has a 
profile which is peaked at the IR end-of-space. The result is that there is a remarkable degree of universality
in the results, which allows to translate the bounds on $S$ directly into model-independent 
bounds on the mass of the techni-rho mesons
$M_{\r}\gsim 2-3$ TeV, with the small discrepancies between the results of various collaborations
 mostly coming from the way in which the comparison to the experimental data is done, more than from the underlying dynamics.
Similar results hold in the top-down context, where  electroweak symmetry and symmetry breaking
are introduced in the string-theory construction 
via probe D-branes~\cite{stringS}, which dynamically develop a non-trivial U-shape embedding in the geometry, along the lines 
of the Witten-Sakai-Sugimoto model of the dual to QCD~\cite{SS}. 
We want to question how general this result is, namely whether it is at all possible to
lower the overall scale $\Lambda_0$ of the strong dynamics without conflicting with precision electroweak physics.
We do so by building a toy model in which 
electroweak symmetry breaking is triggered by a bulk scalar such that its effects are localized at a point in the bulk far away from the IR.
As we will see, this yields a parametric suppression of the $S$ parameter (as was found in ~\cite{FPV}).
The idea behind this construction is that it might be possible to identify the electro-weak group 
with a subgroup of the large symmetry group of the five-dimensional sigma-model, the bulk gauge bosons
with the spin-1 fields emerging from the KK decomposition of 10 dimensional supergravity down to five dimensions, and
 the bulk Higgs with  one of the scalars of the sigma-model.
Whether this is possible in a rigorous string-theory model is a completely open question. Notice also that one would need to check that it is
possible to weakly gauge the global symmetry dual to the bulk symmetry, which requires understanding 
in detail how to perform holographic remormalization~\cite{HR} in this case, in order to
retain in the spectrum otherwise non-normalizable modes.

Finally, we perform a phenomenological study of the experimental data, 
with a three-parameter $\chi^2$ analysis. We confirm  the result of other collaborations that 
 the general dilaton hypothesis is marginally favored over the SM Higgs  one,
after combining ATLAS, CMS and TeVatron measurements.
Most importantly, we want to compare to the predictions we have for the  decay constant
of the holographic techni-dilaton.
We highlight the fact that the main piece of experimental information that 
seems to be in serious contradiction with the dilaton hypothesis is
the TeVatron study of the new scalar produced in association with an
electroweak gauge boson and decaying into a $b\bar{b}$-pair,
while all other measurements either favor the techni-dilaton, or are affected by such large uncertainties
that their weight in the $\chi^2$ is not very important.
In general, because  the holographic techni-dilaton has a decay constant 
of ${\cal O} (1\, {\rm TeV})$,  this would suppress 
all the production mechanisms except for the gluon-gluon-fusion.
The open question in this case is mostly an experimental one: is it possible, with more
data and a better understanding of the backgrounds and of the production mechanisms,
to firmly establish whether this is the case?
At present, the situation is far from clear: besides the fact that the error bars are quite large,
and hence everything is compatible with everything else (at the $3\sigma$ level), 
there appears to be some tension within the data itself, with several independent measurements 
of the same physical quantity yielding contrasting results.
With the anticipated future program of the LHC experiments, it should be possible to settle this question.

In summary, we start from the observation that the simplest models of holographic techni-dilaton
yield two phenomenological predictions. Namely, the decay to photons 
can be enhanced, while the decay constant is significantly larger than the electroweak scale $F\gg v_W$,
which implies that all the production mechanisms except gluon-gluon fusion are suppressed.
From a theoretical perspective, we ask whether the decay constant can be lowered, either by changing the geometry of the model,
or by changing the mechanism of electroweak symmetry breaking. From a phenomenological viewpoint,
 we question whether there is
in fact a clear indication that such suppression is present (or not) in the experimental data.
We then identify clear strategies in order to answer these three questions in the near future.

\section{Formalism}

In this section we summarize some general results and notation,
borrowing partial results from the literature~\cite{BHM,EP,LP,EPdilaton}.
We use the formalism of five-dimensional sigma-models coupled to gravity 
in order to define the background and the formalism of gauge-invariant 
perturbations of the metric and sigma-model scalars in order to
compute the spectrum of scalar composite states and the decay constant 
of the holographic techni-dilaton.

\subsection{Geometry}

The metric of the five-dimensional space-time is, in full generality, given by
\begin{equation}\label{metric}
ds^2=e^{2A(r)}\eta_{\mu\nu}dx^\mu\,dx^\nu+dr^2,
\end{equation}
where $r$ is the extra dimension, $A(r)$ is the warp factor, capital Roman indices span 
$M,N=0,1,2,3,4$, lower case Greek indices span $\mu,\nu=0,1,2,3$ and $\eta_{\mu\nu}$ has signature $(-,+,+,+)$. 
$A(r)$ is independent of the space-time directions $x^\mu$ and, if $A(r)$ is linear, we recover 
an AdS space. To this we add two $3+1$ dimensional boundaries, an IR boundary at $r=r_1$ 
and a UV boundary at $r=r_2$. 
 The UV boundary acts as a regulator and the limit $r_2\rightarrow\infty$ 
 should be taken in subsequent calculations.  The IR boundary is necessary for the numerical calculations,
 and will be removed by extrapolating the results for $r_1\rightarrow 0$.

For simplicity, we write here explicitly the relevant equations for the case in which only one bulk scalar $\Phi$
controls the geometry, the generalization to more than one scalar being  in the literature~\cite{EP}.
 The action for the bulk scalar coupled to gravity is~\cite{EP}
\begin{eqnarray}
\mathcal{S}&=&\int \di^4x \int_{r_1}^{r_2}\di r\left\{ \sqrt{-g}\left(\frac{R}{4}+\mathcal{L}_5\right)+\sqrt{-\tilde{g}}\delta(r-r_1)\left(-\frac{K}{2}+\mathcal{L}_1\right)\nonumber\right.\\ &&\left.-\sqrt{-\tilde{g}}\delta(r-r_2)\left(-\frac{K}{2}+\mathcal{L}_2\right)\right\}.
\end{eqnarray}
where $R$ is the Ricci scalar and $K$ is the extrinsic curvature of the boundary hyper-surface~\cite{EP}.
The sigma-model Lagrangians are
\begin{eqnarray}
\mathcal{L}_5&=&-\frac{1}{2}g^{MN}\partial_M\Phi\partial_N\Phi-V(\Phi)\,,\,\,\,\,
\mathcal{L}_1\,=\,-\lambda_1(\Phi)\,,\,\,\,\,
\mathcal{L}_2\,=\,-\lambda_2(\Phi)\,,
\end{eqnarray}
where $V(\Phi)$ is a bulk potential and the $\lambda_i(\Phi)$ are localized potentials, living on the 4D boundaries.

If the potential $V$ is such that it can be written in terms of a superpotential as
\begin{equation}
V=\frac{1}{2}(\partial_\Phi W)^2-\frac{4}{3}W^2,
\end{equation}
then also the $\lambda_i$ can be expanded in terms of this superpotential
\begin{equation}
\lambda_i=W(\Phi_i)+\partial_\Phi W|_{\Phi_i}(\Phi-\Phi_i)+....
\end{equation}
and  the bulk equations are satisfied by any solutions
of the first-order equations
\begin{equation}\label{eqn:Aprime}
A^\prime=-\frac{2}{3}W,
\end{equation}
and
\begin{equation}\label{Phiprime}
{\Phi}^{\prime}=\partial_\Phi W.
\end{equation}

\subsection{Scalar spectrum and decay constant}

To calculate the mass spectrum of scalar bound states present in this model, one needs 
to consider fluctuations about the classical background. 
Mixing occurs between the fluctuations of $\Phi$ and the scalar fluctuations of the (five-dimensional) 
metric. These fluctuations contain both physical and unphysical degrees of freedom. 
Thankfully, a simple algorithm exists for calculating the 
equations of motion~\cite{BHM} for the gauge-invariant, physical combinations of such fluctuations. 
Applying this process eventually yields the gauge-invariant equations of motion~\cite{BHM}~\footnote{Here and in the following we  define the 
 four-momentum $q^2=-\eta_{\mu\n}q^{\mu}q^{\nu}$.
This seemingly bizarre choice, and a related redefinition of the signs in the vacuum polarization tensors
which we discuss later in the paper,
have the function of facilitating the comparison with the literature. When discussing 
string-theory and supergravity sigma-models, it is customary to adopt the convention in which
the metric has signature $\{-,+,+,+\}$, as we did here. However, when discussing the phenomenology
of Kaluza-Klein theories and of precision electroweak measurements, it is customary to adopt the
convention where the metric has signature $\{+,-,-,-\}$. With this change of sign in $q^2$,
all the equations agree with those in the literature. In particular, the on-shell condition for a particle of mass
$m$ is $q^2=m^2>0$.}
\begin{equation}\label{gieom}
\left(\left(\partial_r+N-\frac{8}{3}W\right)\left(\partial_r-N\right)+e^{-2A}q^2\right)\mathfrak{a}=0,
\end{equation}
and boundary conditions~\cite{EP}
\begin{equation}\label{eqn:nbc}
\left.\frac{}{}\left(\frac{e^{2A}}{q^2}\frac{(W_\phi)^2}{W}\right)(\partial_r-N)\mathfrak{a}\right|_{r_i}=\left.\frac{}{}\mathfrak{a}\right|_{r_i},
\end{equation}
where $N=W_{\Phi\Phi}-\frac{(W_\Phi)^2}{W}$ and $\mathfrak{a}$ is the gauge-invariant variable, defined by
\begin{equation}
\mathfrak{a}=\varphi+\frac{W_\Phi}{4W}h\,.
\end{equation}
$\varphi$ is the fluctuation of the scalar and $h$ is one of the fluctuations of the metric
(see~\cite{BHM,EP} for details).

In order to compute the decay constant, we use the fact that the original fluctuation of the 
four-dimensional $h$ has the following equation of motion:
\beqs
h&=&4e^{2A}\Box^{-1}W_{\Phi}\left(\partial_{r}-N\right)\mathfrak{a}\,.
\eeqs
Making use of partial results from~\cite{LP}, the decay constant of the dilaton F is given by~\footnote{Notice that we reabsorbed a numerical factor into the definition of $F$, which is now defined so that it can be 
compared directly with the Standard Model $v_W$. In the notation of~\cite{LP} we have $F=4f$. Notice also that we conventionally choose the normalizations of the fluctuations 
so that $F>0$.}
\beqs
\frac{1}{F}&\equiv&-\frac{v_{d\gamma\gamma}\sqrt{N}}{\Lambda_0}\,,
\eeqs
where $\Lambda_0$ is the overall mass  scale, 
while
\beqs
v_{d\gamma\gamma}&=&-\frac{1}{3}\int_{r_1}^{r_2} \di r\,h\,,\\
\frac{1}{N}&=&\frac{1}{12}\int_{r_1}^{r_2} \di r\,e^{2A} h^2\,.
\eeqs

All of these equations generalize to the case in which more than one scalar appear in the
sigma-model action. We do not report here the details, which can be found in~\cite{EP}.

\section{The decay constant and the five-dimensional geometry}

The study in~\cite{LP} shows that for  a sigma-model with cubic superpotential, such that 
the solution is a kink, one finds $F/\Lambda_0 \simeq \sqrt{3/2}$, where $\Lambda_0$ is the 
mass scale of the theory. Remarkably, this result is almost independent of the parameters of the model,
and agrees with the result obtained in models where the superpotential is quadratic (which implement and generalize the 
Goldberger-Wise mechanism), and in general in models where the basic geometry is always very close to 
AdS~\cite{dilaton5D}.

In order to gauge how general this is, we perform here the exercise of computing the 
decay constant for a completely different class of models, in which the end-of-space in the IR
arises dynamically due to the bulk profile of the scalar, rather than being put in by hand 
as a hard IR cut-off. This means that near the IR several background functions diverge,
and hence there is no obvious sense in which the deviation from AdS is parametrically small.

\subsection{A class of toy models}

\begin{figure}[t]
\begin{center}
\begin{picture}(500,140)
\put(20,10){\includegraphics[height=4.5cm]{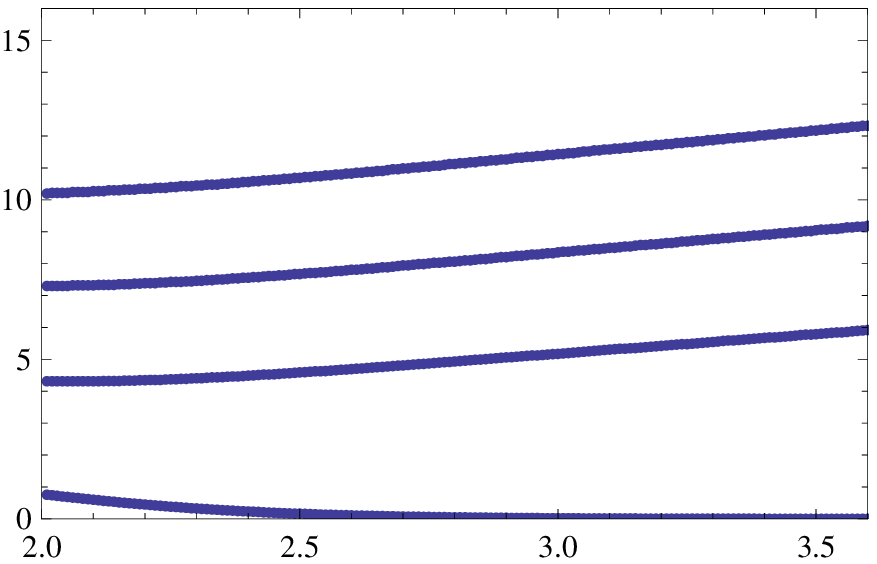}}
\put(270,10){\includegraphics[height=4.5cm]{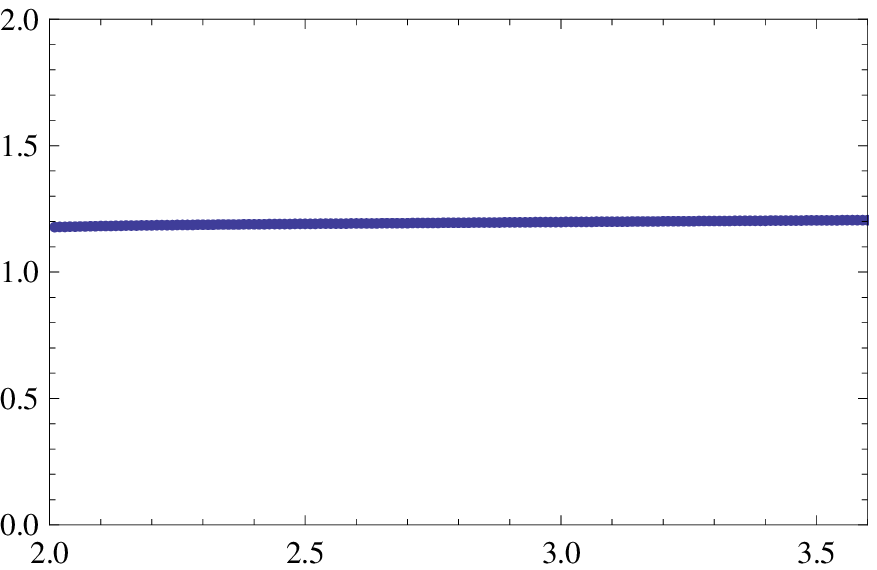}}
\put(2,120){$\frac{m}{\Lambda_0}$}
\put(255,120){$\frac{F}{\Lambda_0}$}
\put(210,6){$\Delta$}
\put(455,6){$\Delta$}
\end{picture} 
\caption{Numerical study of the mass $m$ of the four lightest scalar 
states and  the decay constant $F$ (in units of $\Lambda_0$)
of the lightest of them (the holographic pseudo-dilaton), for the model defined by Eq.~(\ref{Eq:sinh}). The plots are obtained  by varying $\Delta$,
while keeping $r_1=0.01$ and $r_2=5$.}
\label{Fig:study}
\end{center}
\end{figure}

This exercise completes the calculation of the decay constant for all the examples
of simple five-dimensional  models with one bulk scalar
yielding  a dilaton,  the spectrum of which we studied in~\cite{EP}.
We consider the following superpotential, for a scalar with canonical kinetic term
\beqs
W&=&-\frac{4}{3}\left(1+\cosh2\sqrt{\frac{\Delta}{3}}\Phi\right)\,,
\label{Eq:sinh}
\eeqs
which determines the background
\beqs
\bar{\Phi}&=&\sqrt{\frac{3}{\Delta}}\arctanh e^{-\Delta r}\,,\\
A&=&\frac{1}{2\Delta}\ln \left[\frac{}{}-1+e^{2\Delta r}\right]\,.
\eeqs
Notice that we set two integration constants in such a way that the singularity is at $r\rightarrow 0$
and that the far-UV geometry is AdS with unit curvature.
For large $r$ this is approximately AdS, which means that
the constant $\Delta$ can be associated with the scaling dimension of the operator
of relevance to the dual theory.
We consider in this subsection only choices of  $\Delta>2$, so that we know that there is going to be a massless
dilaton in the spectrum (because this means that in the dual field-theory language an operator with dimension larger than 2 is developing a VEV).
We checked this explicitly in~\cite{EP}, where we performed the complete calculation of the spectrum 
for the special case $\Delta=3$ and found a massless state, and for $\Delta=1$, in which case no massless state exists.

In Fig.~\ref{Fig:study} we show the scalar mass spectrum $m$ and the decay constant $F$,
computed with $r_2$ and $r_1$ kept fixed, as a function of the only parameter having a physical meaning, namely the dimension $\Delta$.
Notice that because $r_2$ is finite, the mass of the dilaton (the lightest scalar) is finite: in the limit  $r_2\rightarrow +\infty$
this mass vanishes, which we checked numerically.
The remarkable result is that, in spite of the fact that a divergence is present in the deep IR of the geometry,
the decay constant of the holographic techni-dilaton agrees with the result one gets in many other models,
and does not depend significantly on the parameter $\Delta$. 
On the one hand, this is a good result: it exemplifies how many fundamental quantities show a universal 
behavior, in the sense that the precise details of the extra-dimension scenario are not important.
In a sense, this is the very basis of the fact that making this type of calculations in toy models is actually useful:
some important  results have a general validity, and do not depend on (unknown) details of the exact gravity dual 
to a given field theory of interest.

However, at this point it is still not clear what $\Lambda_0$, the overall mass scale, is, and hence this conclusion is somewhat premature.
We will address this shortly when we ask this question by introducing probe spin-1 fields in the geometry.
Also, the presence of an exactly massless scalar in the spectrum is hardly a welcome feature in a phenomenological model
of relevance to LHC physics.
The reason why such a state is there is clear from the field theory point of view: this is the dual description of a conformal field theory in
which the only non trivial feature in the dynamics is the arising of a condensate for one of the CFT operators.
As such, there is only spontaneous breaking of scale invariance, which results in the presence of an exact dilaton.
But then it is also clear that one should not take this result too literally, because there is no reason why the CFT should yield a
condensate in the first place. In order for this to occur one generically has to add some deformation to the theory, either via higher-order operators
(which at the technical level translates into keeping a finite UV cutoff $r_2$) or by adding mass deformations.
The latter is what we will do later, by discussing the GPPZ model. 
In practice, models with superpotential such as in Eq.~(\ref{Eq:sinh}) should be thought of as a leading-order approximation to more
complete models.
Remarkably, this fact also  exemplifies a well-known counter-example to the universality argument we exposed in the previous 
paragraph: while the calculation of the decay constant is very robust and model-independent, the calculation of
the mass of the techni-dilaton is not, and it requires to know in detail the whole dynamics.
In particular, computing the mass of the techni-dilaton from first  principles is a significantly more difficult and model-dependent 
problem than computing  its decay constant or (as we are going to see) the mass spectrum of the techni-rho mesons,
and it can be done in a fully reliable way only by specifying the full gravity dual (obtained from 10-dimensional string theory).

\subsubsection{Electroweak symmetry and symmetry breaking}

We build a model of electroweak symmetry breaking by making use of the backgrounds
of this section. This is per se an interesting exercise, which generalizes the studies in~\cite{AdSTC}
to a model in which the IR end-of-space is due to the bulk dynamics, as opposed to being 
put in by hand.
For simplicity, we 
 model the SM gauge interactions by dispensing with the charged gauge bosons.
 We introduce a bulk $U(1)_L\times U(1)_R$
gauge symmetry and chiral symmetry breaking 
$U(1)_L\times U(1)_R\rightarrow U(1)_V$ occurs due to different  IR 
boundary conditions for vector and axial-vector fields. The probe
action is
\beqs\label{eqn:gaugeaction}
\mathcal{S}_{gauge}&=&-\frac{1}{4}\int d^4x\int^{r_2}_{r_1}dr\,\left\{ \left(\frac{}{}a(r)-D b(r)\delta(r-r_2)\right)F_{\mu\nu}F^{\mu\nu}
+2b(r)F_{r\mu}F^{r\mu}
\,\nonumber \right.\\
&&\left.\frac{}{} 
-2b(r){\Omega^2}{}A^{\mu} A_\mu\delta(r-r_1)\right\}\,,
\eeqs
where the field strength $F$ stands for the field strength of both the $U(1)_{L,R}$ gauge bosons,
while $A$ is the axial gauge bosons. We have included a VEV $\Omega$ in the IR and a UV-boundary kinetic term $D$.
Notice that the kinetic term $D$ is universal.
The functions $a(r)$ and $b(r)$  parametrise the effect of the metric and the measure,
 and with our choice of coordinates they are given by
\begin{eqnarray}
a(r)&=&1,\\
b(r)&=&e^{2A}.
\end{eqnarray}

The spectrum can be determined from the bulk equation for the function
$\gamma(q^2,r)=\partial_rv(q^2,r)/v(q^2,r)$, where $v(q^2,r)$ is the profile of the gauge bosons in the fifth dimension.
The bulk equation  is given by
\beqs
\partial_r(b(r)\gamma(q^2,r))\,+\,b(r)\gamma(q^2,r)^2+a(r) q^2&=&0\,,
\eeqs
subject to the boundary conditions (in the IR)
\beqs
\gamma_V(q^2,r_1)&=&0\,,\\
\gamma_A(q^2,r_1)&=&\Omega^2\,.
\eeqs
The  axial-vector field $A_{\mu}$ and the 
 vector field $V_{\mu}$ are defined by
\beqs
\label{eq:AVdefinition}
\left\{\begin{array}{ccc}
V^M &=& \sin\theta_W L^M \,+\,\cos\theta_W R^{M}\,\cr
A^M &=& \cos\theta_W L^M \,-\,\sin\theta_W R^{M}\,\cr
\end{array}\right.\,,
\eeqs
with $\theta_W$ the mixing angle of the electro-weak theory.

The  physics is then determined by the UV-boundary action via the 
vacuum polarizations, which are given by
\beqs
\pi(q^2)&=&-{\cal N}\left(q^2 D b(r_2) +b(r_2)\gamma\right)\,,
\eeqs
with 
\beqs
{\cal N}&\equiv&\varepsilon^2\,,\\
D b(r_2)&=&r_2-\frac{1}{\varepsilon^2}\,,
\eeqs
in such a way that the limit $r_2\rightarrow +\infty$ can be taken
yielding a non-trivial, finite result for all the $\pi(q^2)$~\cite{HR}.
The spectrum is given by the zeros of $\pi_V(q^2),\pi_A(q^2)$.

The $\hat{S}$ parameter~\cite{Peskin}, which is the most tightly constrained
by precision electroweak physics, can be computed as~\cite{Barbieri}
\beqs
\hat{S}&=&\cos^2\theta_W\left(\frac{}{}\pi^{\prime}_V(0)-\pi^{\prime}_A(0)\right)\,\lsim\,0.003\,,
\eeqs
with the upper bound an indicative $3\sigma$ bound obtained by extrapolating from the electroweak fits~\cite{Barbieri}.
In order to compute $\hat{S}$, it useful to expand $\gamma(q^2, r)$ in powers of $q^2$,
so that $\gamma(q^2,r)=\gamma_0+q^2\gamma_1\,+\,\cdots$ 
One then finds that the equations satisfied by $\gamma_i$ are
\beqs
\partial_r\left(b(r)\gamma_0(r)\right)+b(r)\gamma_0(r)^2&=&0\,,\\
\partial_r\left(b(r)\gamma_1(r)\right)+2b(r)\gamma_0(r)\gamma_1(r)+a(r)&=&0\,,
\eeqs
with the IR-bounadry conditions
\beqs
\gamma_0(r_1)&=&\Omega^2\,,\\
\gamma_1(r_1)&=&0\,,
\eeqs
where it is understood that $\Omega=0$ for the vector case.

The equations can be integrated formally, to yield
\beqs
\gamma_0(r)&=&\frac{1}{b(r)\left(\frac{1}{b(r_1)\Omega^2}+\int_{r_1}^r\di \r \frac{1}{b(\r)}\right)}\,,\\
\gamma_1(r)&=&-\frac{1}{b(r)}e^{-2\int_{r_1}^r d \r \gamma_0(\r)}\,
\int_{r_1}^r\di \r\, a(\r) e^{2\int_{r_1}^{\r} d \sigma \gamma_0(\sigma)}\,.
\eeqs

In the case $\Omega^2\rightarrow +\infty$, it is convenient to define $X\equiv1/\gamma$,
for which the bulk equation is
\beqs
-b(r)^2\partial_r\left(\frac{X(r)}{b(r)}\right)+b(r)+q^2a(r)X(r)^2&=&0\,,
\eeqs
subject to the boundary condition 
\beqs
X(q^2,r_1)&=&0\,.
\eeqs
Expanding for small $q^2$ we find $X=X_0+q^2 X_1 +\cdots$, with
\beqs
X_0(r)&=&b(r)\int_{r_1}^r\di \r \frac{1}{b(\r)}\,,\\
X_1(r)&=&b(r)\int_{r_1}^r\di \r \frac{a(\r)X_0(\r)^2}{b(\r)^2}\,.
\eeqs

\subsubsection{Results}
We consider  here only  the case $\Omega^2\rightarrow+\infty$.
We find that for this specific model
\beqs
\gamma_0^V&=&0\,,\\
\gamma_1^V&=&-r(-1+e^{2\Delta r})^{-\frac{1}{\Delta}}\,,\\
X_0^A&=&(-1+e^{2\Delta r})^{\frac{1}{\Delta}}\left(\frac{\pi  \csc \left(\frac{\pi }{\Delta}\right)-B_{e^{-2 \Delta
   r}}\left(\frac{1}{\Delta},\frac{\Delta-1}{\Delta}\right)}{2
   \Delta}\right)\,,\\
   &\simeq&(-1+e^{2\Delta r})^{\frac{1}{\Delta}}\left(\frac{\pi  \csc \left(\frac{\pi }{\Delta}\right)}{2
   \Delta}-\frac{e^{-2 r}}{2}\right)\,,\\
   &\simeq & b(r)\, \frac{\pi}{2\Delta\sin\frac{\pi}{\Delta}}\,,\\
X_1^A
   &\simeq&
   \frac{e^{-4 r} \left(-1+e^{2 \Delta r}\right)^{\frac{1}{\Delta}}
   \left(-\Delta^2-8 e^{3 r} \pi  \csc \left(\frac{\pi
   }{\Delta}\right) \sinh (r) \Delta+e^{4 r} \left(\Delta^2+4
   \pi ^2 r \csc ^2\left(\frac{\pi }{\Delta}\right)\right)\right)}{16
   \Delta^2}\,,\\
   &\simeq&b(r)\,
   \frac{1}{16\Delta^2}\left(\Delta^2+4
   \pi ^2 r \csc ^2\left(\frac{\pi }{\Delta}\right)-4\pi\Delta\csc\left(\frac{\pi}{\Delta}\right)\right)\,,
\eeqs
where the approximate expressions are very accurate at large-$r$ and for large-$\Delta$.

Having put all of this in place, we see that
\beqs
\pi_V^{\prime}(0)&=&1\,,\\
\pi_A^{\prime}(0)&=&-\varepsilon^2\left(r_2-\frac{1}{\varepsilon^2}-b(r_2)\frac{X_1^A(r_2)}{X_0^A(r_2)^2}\right)\,\\
   &=&1+\varepsilon^2\left(\frac{\Delta^2 \sin ^2\left(\frac{\pi }{\Delta}\right)-4 \Delta
   \pi  \sin \left(\frac{\pi }{\Delta}\right)}{4 \pi ^2}\right)\,.
\eeqs
Finally, the $\hat{S}$ parameter is
\beqs
\label{eq:ShatGPPZapprox}
\hat{S}&=&\cos^2\theta_W\left(\frac{}{}\pi^{\prime}_V(0)-\pi^{\prime}_A(0)\right)\,=\,
\cos^2\theta_W\,\varepsilon^2\left(\frac{-\Delta^2 \sin ^2\left(\frac{\pi }{\Delta}\right)+4 \Delta
   \pi  \sin \left(\frac{\pi }{\Delta}\right)}{4 \pi ^2}\right)\,.
\eeqs
Notice that for $\Delta\rightarrow+\infty$ one finds $\hat{S}\rightarrow \frac{3}{4}\varepsilon^2\cos^2\theta_W$,
which is the AdS result with IR Dirichelet boundary conditions~\cite{AdSTC,EPdilaton}.
The reason for this is that by taking $\Delta\rightarrow +\infty$ in practice the background becomes identical
to the AdS with an IR hard-wall at $r=0$.

As a side remark, notice also that our  result for $\hat{S}$ vanishes for $\Delta=1$. The reason for this is that in this case
the bulk dynamics is controlled by a free scalar in the dual theory. However, 
there is no light dilaton in the spectrum for $\Delta<2$.
For values of $\Delta>2$
the result is very close to the asymptotic one, with a modest reduction for values close to $\Delta=2$,
for which $\hat{S}\rightarrow \cos^2\theta_W\varepsilon^2(2\pi-1)/\pi^2$.

\begin{figure}[t]
\begin{center}
\begin{picture}(250,140)
\put(15,10){\includegraphics[height=4.5cm]{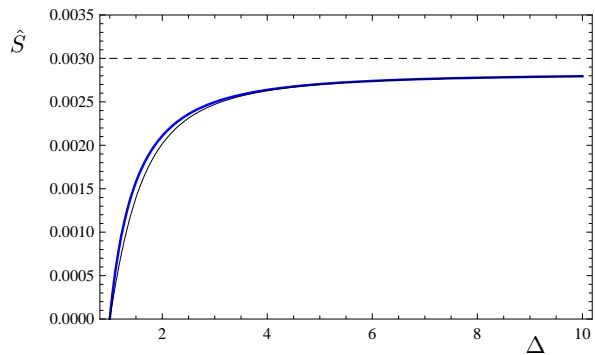}}
\put(0,120){$\hat S$}
\put(195,6){$\Delta$}
\end{picture} 
\caption{$\hat S$ as a function of $\Delta$ for the model defined by Eq.~(\ref{Eq:sinh}), and $\varepsilon=0.07$: the numerical result (blue), the approximation Eq.~\eqref{eq:ShatGPPZapprox} (black), and the experimental bound (dashed black).}
\label{Fig:GPPZSvsDelta}
\end{center}
\end{figure}

The dependence on the free parameter $\varepsilon$ needs to be translated into a dependence on physical quantities.
We do so by computing the mass of the $Z$ bosons,
for which we can use the approximation
\beqs
\pi^A(0)&=&-{\cal N} b(r_2) \frac{1}{X_0(r_2)}\,\simeq\,-\varepsilon^2 \frac{2\Delta \sin\frac{\pi}{\Delta}}{\pi}\,.
\eeqs
and hence
\beqs
\frac{M_Z^2}{\Lambda_0^2}&=&\varepsilon^2 \frac{2\Delta \sin\frac{\pi}{\Delta}}{\pi}\,.
\eeqs
This means that for $\Delta\simeq 2$
\beqs
\hat{S}&\simeq&\cos^2\theta_W \frac{M_Z^2}{\Lambda_0^2}\frac{2\pi-1}{4\pi}\,,
\eeqs
while for $\Delta\rightarrow +\infty$
\beqs
\hat{S}&\simeq&\frac{3}{8}\cos^2\theta_W \frac{M_Z^2}{\Lambda_0^2}\,,
\eeqs
which implies a very modest $12\%$ difference between these two extrema,
and hence for all the range  of interest the bound on $\hat{S}$ is $\Lambda_0\simeq 900$ GeV. We verified numerically that all the approximations are accurate (see Fig.~\ref{Fig:GPPZSvsDelta}).

\begin{figure}[t]
\begin{center}
\begin{picture}(250,140)
\put(15,10){\includegraphics[height=4.5cm]{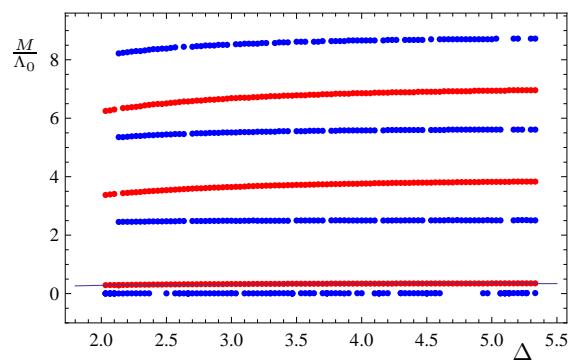}}
\put(0,120){$\frac{M}{\Lambda_0}$}
\put(190,6){$\Delta$}
\end{picture} 
\caption{Numerical study of the mass $M$  (in units of $\Lambda_0$)
for the first few spin-1 states of the model based on the background coming from Eq.~(\ref{Eq:sinh}). 
The plots are obtained  by varying $\Delta$,
while keeping $\varepsilon=0.25$, $r_1=0.001$ and $r_2=10$. The axial-vector points are for $\Omega^2=550$,
which is a very good approximation of $\Omega\rightarrow +\infty$.
The blue dots are numerical results for the spectrum of vector states, while the red dots are the
axial-vector states. The line is the approximation for $M_Z$ in the text.}
\label{Fig:spin1}
\end{center}
\end{figure}

Furthermore, the spectrum of heavy techni-rho mesons is approximately given by the zeros of the Bessel function $J_0(q)$,
so that the lightest such state is approximately $M_{\r}\simeq 2.5 \Lambda_0$.
All of this meaning that we obtain pretty much the same results as for the pure AdS background, 
in particular not only $F\simeq 1.2 \Lambda_0$, but also $M_{\r}\simeq 2.5 \Lambda_0$
are almost independent of $\Delta$ and agree with the AdS.
And also the bounds from the precision parameter $\hat{S}$ are practically unchanged, with $\Lambda_0\gsim 900$ GeV~\cite{AdSTC, EPdilaton, LP}. These results are obtained with a set of approximations that we checked by solving numerically the exact equations.
In Fig.~\ref{Fig:spin1} we show the spectra of vector and axial-vector, neutral states.
The approximation we wrote is quite accurate for $M_Z$, and particularly when
 $\Delta\rightarrow +\infty$ the whole spectrum becomes  indistinguishable from the
case of AdS background with a hard-wall IR cutoff.

In this study, we considered only the contribution to the S-parameter coming from the direct (tree-level) mixing of the electroweak gauge bosons with (all) the spin-1 composite states, while ignoring all possible indirect (loop-induced) effects coming from any other composite states. There are two reasons for doing so. First of all, this is by far the dominant contribution, which for masses of the techni-rho mesons of a few hundred GeV would exceed by orders of magnitude the experimental bounds. And indeed we find a quite strong lower bound on the masses of such spin-1 states, in the multi-TeV range. Even more importantly, this is the only calculable contribution. For example, the contribution from the dilaton itself cannot be computed, but can only be {\it estimated}. Such an estimate shows that the scalar sector yields a contribution that is at most of the same order of magnitude as the experimental uncertainty on the S-parameter itself, and hence we can neglect this contribution for the purposes of this paper.

\subsection{The GPPZ model}

As we explained, for $\Delta>2$ there is an exactly massless dilaton (for $r_2 \rightarrow +\infty$) in the toy models of this section, which is not
phenomenologically acceptable, and is due to the fact that we are assuming the existence of a VEV in an exact CFT. The way around this problem is to extend the model,
by assuming the existence of a second scalar parameterizing the effects of a deformation of the dual field theory by the insertion
of a coupling for a relevant operator.
We do so by borrowing a specific string-theory model, the GPPZ model~\cite{GPPZ}.
We hence repeat the calculation of the scalar spectrum for this more realistic and better motivated 
model, discussing in detail the fate of the very lightest state, and complementing the existing 
literature on the subject~\cite{GPPZspectrum}.

\begin{figure}[t]
\begin{center}
\begin{picture}(500,140)
\put(20,10){\includegraphics[height=4.5cm]{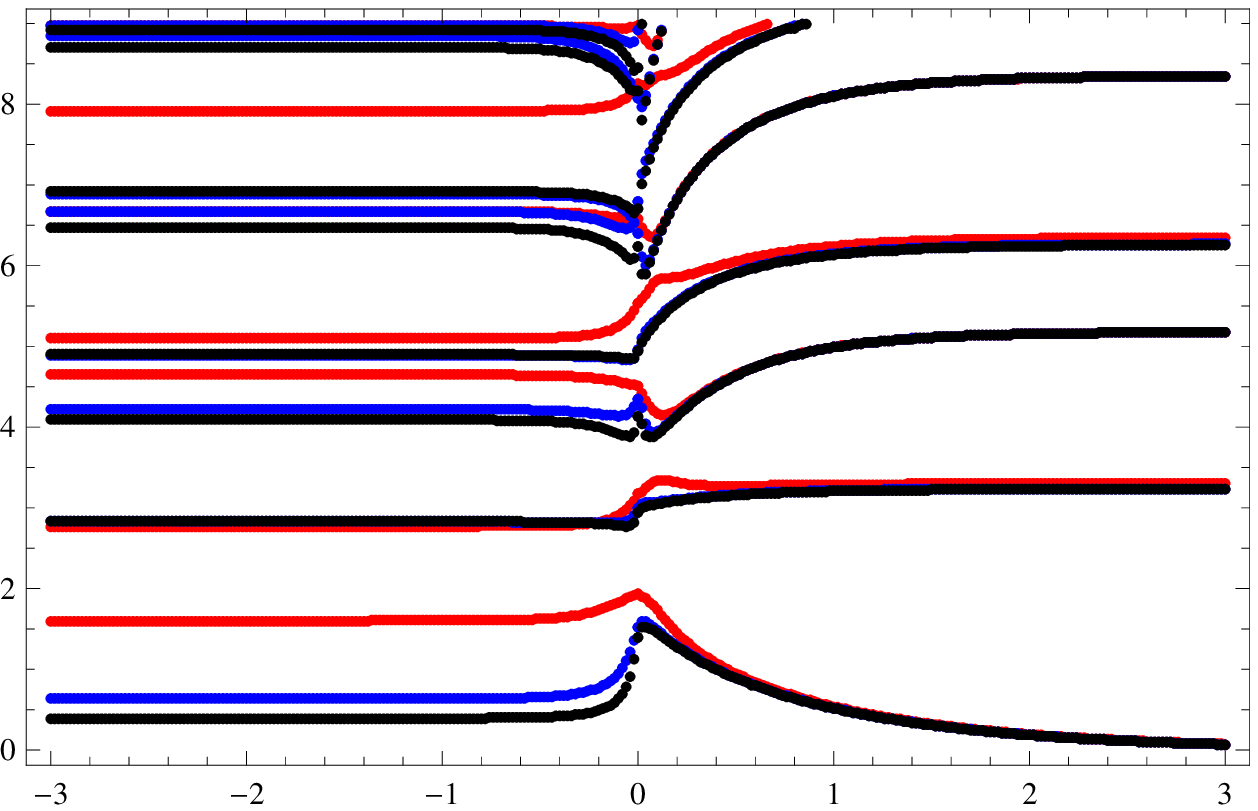}}
\put(270,10){\includegraphics[height=4.5cm]{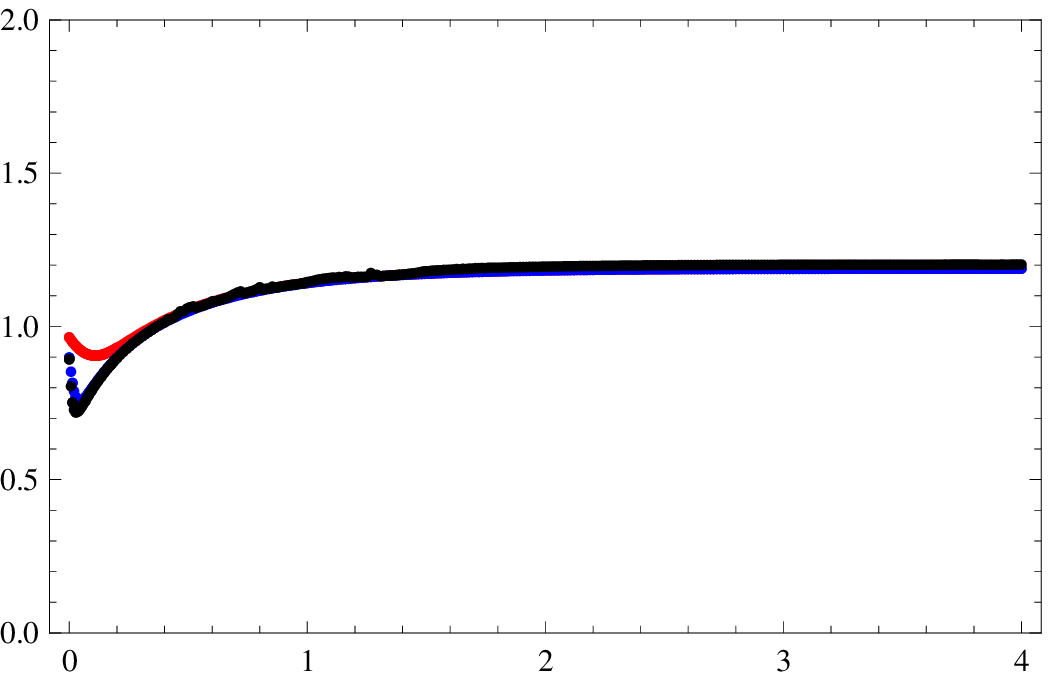}}
\put(2,120){$\frac{M}{\Lambda_0}$}
\put(255,120){$\frac{F}{\Lambda_0}$}
\put(188,6){$c_1-c_2$}
\put(435,6){$c_1-c_2$}
\end{picture} 
\caption{Numerical study of the mass spectrum of scalar composite states and the decay constant $F$ of the holographic techni-dilaton for the GPPZ model. The calculations have been performed for UV cutoff $r_2=10$. The three colors correspond to different choices of IR cutoff, defined through the condition $A(r_1) = -2.5$ (black), $A(r_1) = -2$ (blue), and $A(r_1) = -1$ (red). This means that the black points are obtained for values that are closest to the IR end-of-space. Shown in the left panel is the mass $M$ of the scalar states, as a function of $c_1-c_2$: when $c_1-c_2>0$ the end-of-space is
due to the behavior of $\sigma$, and vice-versa due the behavior of $m$ when $c_1-c_2<0$. The right panel shows the decay constant $F$ (in units of $\Lambda_0$) as a function of $c_1-c_2$ with $c_1=0$.}
\label{Fig:GPPZ}
\end{center}
\end{figure}

Following the notation of Pilch and Warner~\cite{PW}, the superpotential of GPPZ is
\beqs
W&=&-\frac{3}{4}\left(\cosh 2\sigma +\cosh \frac{2m}{\sqrt{3}}\right)\,,
\eeqs
with canonical kinetic terms for $\sigma$ and $m$.
The former is interpreted within the context of deformations of ${\cal N}=4$~SYM
as the gaugino condensate ($\Delta=3$), the latter as a mass term
for the matter fermions ($\Delta=1$).

If one truncates further, by setting either $m=0$ or $\sigma=0$, one ends up with two
examples studied in~\cite{EP}, which we have already mentioned earlier. 
In the former case, the spectrum of the resulting truncated system containing only $\sigma$ 
yields a parametrically light state (in connection with the fact that $\Delta=3$ in this case), 
while in the latter case all the composite scalars are massive (because $\Delta=1$).
Here we want to understand what happens in the more general case, where both scalars are present.

The solution of the $(m,\sigma)$ system is discussed in~\cite{PW} and in~\cite{GPPZ}.
It is given by
\beqs
\sigma&=&\arctanh \left(e^{-3r+3c_1}\right)\,\simeq\,e^{3c_1} \xi^3\,,\\
m&=&\sqrt{3}\arctanh\left(e^{-r+c_2}\right)\,\simeq\,\sqrt{3}e^{c_2} \xi\,,\\
e^{2A}&=& e^{-2r}\left(-1+e^{6(r-c_1)}\right)^{1/3}\left(-1+e^{2(r-c_2)}\right)e^{2c_1+2c_2}\,\simeq\,e^{2r}\,,
\eeqs
where for convenience we show also the UV-expansions, with $r=-\log \xi$.
We kept the two integration constants $c_1$ and $c_2$ explicit, while 
setting the integration constant in $A$ so that the UV is asymptotically AdS with unit curvature.

The resulting solution depends on the $c_1$ and $c_2$ integration constants.
This can be interpreted in terms of the UV-value of the dimension-1 coupling and dimension-3 VEV,
or they can be more directly related to the singular behavior of the metric in the IR.
Indeed, $e^{2A}$ vanishes when $r\rightarrow c_{1,2}$.
Three possibilities emerge, depending on $c_1-c_2$.
Unfortunately, according to~\cite{PW}, who discuss the full lift to 10-dimensions,
all of them yield singular 10-dimensional backgrounds, in various different ways,
but we are not concerned with this issue in the present discussion.

If $0=c_1>c_2$, the space ends in a singularity at $r\rightarrow 0$. In this case, one finds
that the singular behavior is dominated by the dimension-3 VEV. In particular, for $c_2\rightarrow -\infty$
one recovers the $\Delta=3$ case of the toy model, and hence one expects the 
presence of a light scalar. The fact that $c_2$ is finite, means that there is explicit breaking of scale invariance,
and hence the dilaton is actually massive.
For $0=c_2>c_1$,  the singular behavior is due to the $\Delta=1$ insertion.
In this case, we do not expect a light dilaton in the spectrum.
In particular, in the limit $c_1\rightarrow -\infty$, one could truncate $\sigma$ completely,
and should recover the fact that there is no light dilaton, as in~\cite{EP}.

We performed the explicit numerical calculation, for a variety of choices of parameters,
and we illustrate the general results in Fig.~\ref{Fig:GPPZ}.
The plot shows the spectrum of scalar states, computed by using very large values of $r_2$,
for which we checked that changing $r_2$ does not result in any appreciable change in the spectrum.
We computed the spectrum for a variety of choices of $c_1-c_2$, always fixing the largest of the two to zero,
so that the IR singularity is always at $r\rightarrow 0$.
We varied also the value of the IR cutoff $r_1$ taking it to be as close as possible to the end-of-space, while at the same time ensuring that the physical scale of the IR stays the same as $c_1-c_2$ is varied.

Let us start to comment on the results starting from $c_1=0$ and $c_2<0$.
In this case, we expect, on the basis of the considerations made earlier on, the presence of a light dilaton
in the spectrum, with mass induced by the presence of the dimension-1 insertion.
For very negative choices of $c_2$, such insertion is small, and hence the dilaton becomes very light.
This is what the numerics shows. Also, the numerical study shows that, as expected, there is no 
problem with the IR cutoff, since the results become independent of $r_1$ provided this is small.
There is also a tower of heavy states, the mass of which does not depend appreciably on $c_2$.

This is the physically interesting regime for the discussion of this paper.
The dynamics yields a parametrically light dilaton, and because of the smallness of $e^{c_2}$
the phenomenology is very similar to the one in the toy model with only one scalar with $\Delta=3$,
except for the crucial difference that the mass of the techni-dilaton does not vanish.
In this case it is of interest to redo the calculation of the decay constant of the dilaton, as a function of $c_2$.
The result is expected to agree with the one-scalar model for $c_2\rightarrow -\infty$.
By just looking at the spectrum, we also see that when $c_2$ comes close to zero, the lightest state is not actually light,
and cannot be a dilaton. In this case, we expect very large mixing terms to emerge, and hence 
the very calculation of the decay constant we perform (which is based on assuming the light state being predominantly $h$),
is not accurate enough. 
We show the numerical results of the calculation in Fig.~\ref{Fig:GPPZ}, confirming that $F/\Lambda_0\simeq 1.2$
also in this case, provided $c_1 - c_2 \gsim 1$.

For completeness, we show in Fig.~\ref{Fig:GPPZ} also the results for the spectrum when $c_1-c_2<0$,
which reveals a surprising result.
Contrary to what happens if one truncates $\sigma=0$ and removes it completely from the 
sigma-model, we find evidence of the presence of a parametrically light state.
This is however {\it not} a dilaton, but rather a spurious state, the origin of which 
lies in the stringy nature of the IR singularity, or, equivalently, in
the moduli space of the dual field theory.

The numerical results support these statements,  on the basis of two observations.
First of all, the mass of this light state does not depend on $c_1$ and $c_2$, i.e. 
it does not depend on the parameters that should control explicit and spontaneous breaking
of scale invariance.
Second, the mass depends dramatically on the value of the IR cutoff $r_1$, and actually vanishes
when the IR cutoff is removed.
Even more: a techni-dilaton would be predominantly the $h$ of the scalar perturbation of the metric,
with small contaminations from the fluctuations of $m$ and $\sigma$.
The very fact that by truncating $\sigma=0$ the light state disappears suggests that it cannot be
a dilaton.

Hence, what is the physical interpretation of such a state?
First of all, we must remind the reader that the introduction of an IR cutoff,
while necessary, is a dangerous tool, because it
 has the potential of actually modifying the long-distance  physics of
the system.
The physical reason for what is actually happening is explained (indirectly) by Pilch and Warner~\cite{PW},
and related to the work of Polchinski and Strassler~\cite{PS}, 
and connects back to the field-theory study of mass deformations of ${\cal N}=4$ SYM.
 The lift to 10 dimensions for $c_1\lsim c_2$
shows a very peculiar structure, in which the metric near the IR singularity takes the form expected
from the presence of  extended objects (NS5 and D5 branes)
 wrapping an internal cycle inside $S^5$ at the origin of the 
radial direction.
This result has an important meaning (and history) in the context of mass deformations of
${\cal N}=4$ SYM, being (non-trivially) related to the vacuum (moduli space) of the dual theory,
which results in a spurious symmetry remaining in the five-dimensional sigma-model,
and a corresponding modulus.
This spurious symmetry is broken explicitly by the  IR cutoff, because the IR cutoff
would hide the 5-branes outside the physical interval of the radial direction.
Removing the IR cutoff makes the corresponding mode exactly massless. But for our present purposes, this is no more than a very interesting curiosity
and we will not talk about it any further, instead restricting our  attention to $c_1-c_2>0$.

\subsection{Summary}

The study in this section was motivated by the fact that the decay constant of the techni-dilaton $F$ and the mass of the lightest 
techni-rho meson $M_{\r}$,
computed in various models based on small deviation from AdS backgrounds in which confinement is due to 
an IR cutoff put in by hand, show a remarkable level of universality.
Hence, we constructed and studied a new class of models of holographic electroweak symmetry breaking, 
in which the end-of-space 
emerges as a dynamical feature, due to a divergence induced by the classical solutions for the 
bulk scalar(s). We performed the study of the spectrum of spin-0 and spin-1 states of the
system, confirming not only that a light holographic techni-dilaton is present in the spectrum,
but also that the results for $F$ and $M_{\r}$ are virtually unaffected by the modifications of the geometry.

One particular model of this class is the GPPZ model, for which we computed the spectrum of spin-0 excitations,
a calculation which had not been carried out in full generality in the literature so far. We found several interesting features 
of the spectrum, which are of general interest.

The main lesson we learned is that even in the presence of a sharp singularity in the deep-IR,
the decay constant of the dilaton is the same as in Randall-Sundum-like models.
We leave open a very interesting question, which we reformulate in three different ways. 
Is this going to hold also for backgrounds that are
very far from AdS over a large region (showing for example hyperscaling violation)?
In other words, what is the dynamical origin of this rigidity of the dilaton decay constant? 
Is it possible at all to produce models in which there exists a parametric separation between the values of the 
dilaton decay constant and the mass of
the lightest techni-rho meson?
We stress that these are completely open questions and leave them for future work.

\section{The electroweak symmetry breaking scale in holography}

In this section we want to argue that it is possible to moderately lower the 
fundamental scale of the model (and hence the numerical value of the techni-dilaton decay constant),
by modifying the way in which electro-weak symmetry breaking is triggered.
We will comment in due time on how realistic this possibility is, here we anticipate only the fact that 
the basic idea allowing to suppress the precision parameter $\hat{S}$
 is not dissimilar from what is discussed in~\cite{FPV}, with the significant difference that
 in the example we are proposing here we have a concrete sigma-model controlling the dynamics,
 while in~\cite{FPV} the background is written by hand, without specifying where it comes from.
 
 Before proposing the example we are interested in, let us summarize the results of~\cite{EPdilaton}
 and~\cite{LP}, which are in general agreement with the literature on higgsless models~\cite{AdSTC}.
 The background is determined by a sigma-model with one scalar $\Phi$ having canonical kinetic term,
  and the superpotential
 \beqs
W&=&-\frac{3}{2}-\frac{\Delta}{2}\Phi^2+\frac{\Delta}{3\Phi_I}\Phi^3\,,
\eeqs
and by the background solution
\beqs
\Phi&=&\frac{\Phi_I}{1+e^{\Delta(r-r_{\ast})}}\,,
\eeqs
where $r_{\ast}$ is an integration constant.
Confinement is modeled crudely, by assuming that the IR cutoff $r_1$ is actually physical,
and keeping it fixed to $r_1=0$
 
In the model in~\cite{EPdilaton,LP}, electroweak symmetry breaking is triggered by assuming that
five-dimensional versions of the electroweak gauge bosons be allowed to propagate in the bulk of the fifth dimension, and that the axial-vector components satisfy either Dirichlet~\cite{EP}
of generalized Neumann~\cite{LP} boundary conditions in the IR.
One finds that in the former case
the mass $M_Z$ of the $Z$ boson, the precision parameter $\hat{S}$, the mass $M_{\r}$ 
of the techni-rho meson and the decay constant $F$ of the techni-dilaton are approximately given by
\beqs
\frac{M_Z^2}{\Lambda_0^2}&\simeq&2\varepsilon^2\,,\\
\hat{S}&\simeq& \frac{3}{4}\varepsilon^2\cos^2\theta_W\,,\\
\frac{M_{\r}}{\Lambda_0}&\simeq&2.5 - 3\,,\\
\frac{F}{\Lambda_0}&\simeq&1.2\,,
\eeqs
where $\theta_W$ is the electroweak mixing angle, and $\varepsilon$ a free parameter which controls
the strength of the electroweak gauge couplings in respect to the self-couplings of the techni-rho mesons.
In~\cite{LP} a new parameter is introduced, which affects the IR boundary conditions for the axial-vector
bosons in such a way that $\Omega\rightarrow +\infty$ reproduces the 
Dirichlet case, while for $\Omega\rightarrow 0$ there is no symmetry-breaking.
In this case, and for small $\Omega$, one has
\beqs
\frac{M_Z^2}{\Lambda_0^2}&\simeq&\varepsilon^2\Omega^2\,,\\
\hat{S}&\simeq& \frac{1}{2}\varepsilon^2\Omega^2\cos^2\theta_W\,,\\
\frac{M_{\r}}{\Lambda_0}&\simeq&2.5 - 3\,,\\
\frac{F}{\Lambda_0}&\simeq&1.2\,.
\eeqs
The main difference between the two cases 
is that by imposing the indicative bound $\hat{S}\lsim 0.003$~\cite{Barbieri},
in the former case one is forced to choose small values of $\varepsilon$ 
(which is related to the coupling of the dilaton to two photons), while in the latter 
case $\varepsilon$ is a free parameter.
We already commented on the fact that the prediction for $F$  is rather robust.
It is also to be noticed that the masses of the techni-rho mesons are only very 
mildly dependent on the parameters, so that ultimately imposing the bound from $\hat{S}$ 
and the numerical value of the mass of the $Z$ boson
fixes the scale $\Lambda_0$ of the theory to be $\Lambda_0\sim 900$ GeV.

As a consequence, one finds that in physical units $F\simeq 1.1$ TeV is a rigid prediction,
which depends only modestly  on the parameters.
This is what we want to challenge here: is it possible to lower $\Lambda_0$?

\subsection{Modelling spin-1 sector and electroweak symmetry breaking from the bulk}

As we did earlier, we restrict only to the $U(1)_L\times U(1)_R$ gauge group, for simplicity.
We start by changing how the spin-1 states are introduced into the model
and how they couple to the source of electro-weak symmetry breaking.
We consider a model in which a bulk Higgs field $H$ is charged under the group corresponding to the axial
$U(1)$ symmetry. We rewrite the complex $H$ as
\beqs
H&=&\frac{1}{\sqrt{2}}\Phi e^{i\theta}\,,
\eeqs
where $\theta$ is the phase. 
Hence, we identify the bulk dynamical field $\Phi$ with the Higgs bulk field that induces electroweak symmetry breaking.
A few words to alert the reader about a subtlety related to gauge fixing. 
By using the canonical kinetic term for $H$,
with the appropriate covariant derivative,
we obtain the kinetic term for $\Phi$ and the mass term for $A^{\mu}$,
which are of interest here.
However, there is also a term which couples at the quadratic order $A^{\mu}$
and $\theta$.
Hence, while the unitary gauge for $V^{M}$ is given by $V_5=0$ (and hence $V_5$
disappears from the action),
the axial $A_5$ and $\theta$ mix. The unitary gauge is defined by setting a 
combination of $\theta$ and $A_5$ to zero, which leaves the orthogonal component 
in the action. Notice that in the case where the symmetry-breaking is due to boundary conditions, this subtlety 
is absent and $A_5=0$ is the unitary gauge.
Yet, in the action we neglect to write
explicitly the whole pseudo-scalar sector originating from this other 
combination of $A_5$ and $\theta$.
The pseudo-scalar sector we are neglecting does not have light (coupled) degrees of freedom, and hence
for the present purposes we are allowed to ignore it.
With all of this, all the interesting physics  is determined only by $\Phi$, which we assume to 
be the one scalar in~\cite{EPdilaton,LP}

We treat the gauge bosons in probe approximations.
Electroweak symmetry breaking is triggered by the fact that the gauge bosons couple to
the bulk scalar, which acts as the 5-dimensional Higgs field.
The probe
action is
\beqs
\mathcal{S}_{gauge}&=&-\frac{1}{4}\int d^4x\int^{r_2}_{r_1}dr\,\left\{ \left(\frac{}{}a(r)-D b(r)\delta(r-r_2)\right)\left(\frac{}{}L_{\mu\nu}L^{\mu\nu}+R_{\mu\nu}R^{\mu\nu}\right)\right.\nonumber\\
&&\left.
+2b(r)\left(\frac{}{}L_{r\mu}L^{r\mu}+R_{r\mu}R^{r\mu}\right)
+2b(r)\tilde{M}^2(r)A_{\mu}A^{\mu}
\,\nonumber \right\}\,,
\eeqs
where we have included  a UV-boundary kinetic term $D$,
and we represent both left and right handed groups by the field-strength tensors.
The mass term $\tilde{M}^2$ is given by
\beqs
\tilde{M}^2&=&\frac{1}{2}\left(g_L^2+g_R^2\right) \Phi^2\,,
\eeqs
where $g_{L,R}$ are the  (five-dimensional) gauge couplings.
The functions $a(r)$ and $b(r)$ are again $a(r)=1$ and
$b(r)=e^{2A(r)}$.

The spectrum can be determined from the bulk equation for the functions
$\gamma_{V,A}(q^2,r)\equiv \partial_r \ln v(q^2,r)$,  where $v(q^2,r)$ is the wave-function of the gauge bosons in the fifth dimension. We write explicitly the bulk equations, which  are  given by
\beqs
\partial_r(b(r)\gamma_V(q^2,r))\,+\,b(r)\gamma_V(q^2,r)^2+a(r) q^2&=&0\,,\\
\partial_r(b(r)\gamma_A(q^2,r))\,+\,b(r)\gamma_A(q^2,r)^2+a(r) q^2-b(r)\tilde{M}^2(r)&=&0\,.
\eeqs
subject to the boundary conditions (in the IR)
\beqs
\gamma_V(q^2,r_1)&=&0\,,\\
\gamma_A(q^2,r_1)&=&0\,.
\eeqs
where the axial-vector $A^{\mu}$,
 the vectorial $V^{\mu}$ fields are defined as usual by Eq.~\eqref{eq:AVdefinition}.
The  physics is then determined by the UV-boundary action via the 
vacuum polarizations $\pi$ which we already defined earlier.

In this case one finds that the equations satisfied by $\gamma_i$ are
\beqs
\partial_r\left(b(r)\gamma_{V\,0}(r)\right)+b(r)\gamma_{V\,0}(r)^2&=&0\,,\\
\partial_r\left(b(r)\gamma_{A\,0}(r)\right)+b(r)\gamma_{A\,0}(r)^2-b(r)\tilde{M}^2&=&0\,,\\
\partial_r\left(b(r)\gamma_1(r)\right)+2b(r)\gamma_0(r)\gamma_1(r)+a(r)&=&0\,,
\eeqs
where the equation for $\gamma_1$ is formally the same for $V$ and $A$.
The IR-boundary conditions are
\beqs
\gamma_0(r_1)&=&0\,,\\
\gamma_1(r_1)&=&0\,.
\eeqs

The equations for the vectorial part can be integrated, to yield
\beqs
\gamma_{V\,0}(r)&=&0,\\
\gamma_{V\,1}(r)&=&-\frac{1}{b(r)}
\int_{r_1}^r\di \r\, a(\r)\,.
\eeqs
For the axial part, this is more subtle.
While it is certainly the case that
\beqs
\gamma_{A\,1}(r)&=&-\frac{1}{b(r)}e^{-2\int_{r_1}^r d \r \gamma_{A\,0}(\r)}\,
\int_{r_1}^r\di \r\, a(\r) e^{2\int_{r_1}^{\r} d \sigma \gamma_{A\,0}(\sigma)}\,,
\eeqs
solving for $\gamma_{A\,0}$ is non-trivial, but
we can approximate the result.
We first expand $\gamma_{A\,0}$ in powers of a small parameter controlling $\tilde{M}^2$
(for example $\Phi_I$, the IR value of the scalar $\Phi$), and look at the equations for
the components of this expansion. We then realize that the leading order equation 
is the same as for the vector part, and hence vanishes.
At the first non-trivial order in $\tilde{M}$ we have then
\beqs
\gamma_{A\,0}(r)&=&\frac{1}{b(r)}
\int_{r_1}^r\di \r\, b(\r)\tilde{M}^2(\r)\,.
\eeqs

In order to compute the precision parameter $\hat{S}$, we need to compute explicitly also $\gamma_{A\,1}$.
To do so, we expand $\gamma_{A\,1}=\gamma_{V\,1}+\delta \gamma_{A\,1}+\cdots$, where $\delta\gamma_{A\,1}$ appears at the same order in $\tilde{M}^2$ as $\gamma_{A\,0}$.
Doing so, we obtain the equation
\beqs
\partial_r\left(b \delta\gamma_{A\,1}\right)+2b\gamma_{A\,0}\gamma_{V\,1}&=&0\,,
\eeqs
subject to the boundary condition $\delta\gamma_{A\,1}(r_1)=0$.
This yields
\beqs
\label{eq:bdgammaA1}
b(r)\delta\gamma_{A\,1}(r)&=&-\int_{r_1}^r \di \r \, 2b(\r)\gamma_{A\,0} (\r)\gamma_{V\,1}(\r)\,.
\eeqs

\subsection{Results}

We do not rediscuss here the physics of scalar and vectorial spin-1 states, which is identical 
to~\cite{EPdilaton,LP}. We focus instead on the axial-vector part of the spectrum.
We perform the approximate calculations by neglecting the backreaction of the bulk scalar on the bulk metric,
which, provided $\Delta \Phi_I$ is small, means we can approximate $b\simeq e^{2r}$.

In this case, the integral for $\gamma_{A\,0}$ 
can be performed analytically, and at the leading order in $\Phi_I$ we have
\beqs
	b(r)\gamma_{A\,0}(r) &\simeq& \frac{1}{4}\left(g_L^2+g_R^2\right)\Phi_I^2
	\left[ e^{2 r}
	\, _2F_1\left(2,\frac{2}{\Delta};\frac{\Delta+2}{\Delta};-e^{\Delta(r-r_{\ast})}\right) -
	\, _2F_1\left(2,\frac{2}{\Delta};\frac{\Delta+2}{\Delta};-e^{-\Delta r_{\ast}}\right)
\right]\,.
\eeqs
This functional  dependence looks  intimidating, but actually has a  simple shape: it is approximately constant
for $r>r_{\ast}$, and exponentially suppressed when $r<r_{\ast}$.
In practical terms, this means that the symmetry breaking effects encoded in $\gamma_{A\,0}$
are actually localized at $r_{\ast}$: for $r>r_{\ast}$ the exponential factor from $b$ suppresses 
$\gamma_{A\,0}$, while the growth for $r<r_{\ast}$ off-sets the factor of $b(r)$.
In practice this means that by choosing $r_{\ast}$ we have a situation which interpolates
between the pure UV-boundary breaking (for which $\hat{S}=0$) and the pure IR-boundary
breaking, and hence we expect to see a suppression.

Provided $\Delta>2$, we have that for $r_{\ast}\gg 0$
\beqs
\lim_{r\rightarrow +\infty}b(r)\gamma_{A\,0}(r) &\simeq& \frac{1}{2}\left(g_L^2+g_R^2\right)\Phi_I^2 
   \frac{(\Delta-2)\pi e^{2 r_{\ast}}}{\Delta^2\sin\frac{2\pi}{\Delta}} \,.
\eeqs
Notice that because $\pi(q^2)$ is related to $-\varepsilon^2\gamma(q^2)$, this means that,
as long as $\hat{S}$ is small, the mass of the lightest state is
\beqs
M_Z^2&\simeq&
\frac{1}{2}\varepsilon^2\left(g_L^2+g_R^2\right)\Phi_I^2
   \frac{(\Delta-2)\pi e^{2 r_{\ast}}}{\Delta^2\sin\frac{2\pi}{\Delta}}
   \,.
\eeqs

\begin{figure}[t]
\begin{center}
\begin{picture}(250,140)
\put(15,10){\includegraphics[height=4.5cm]{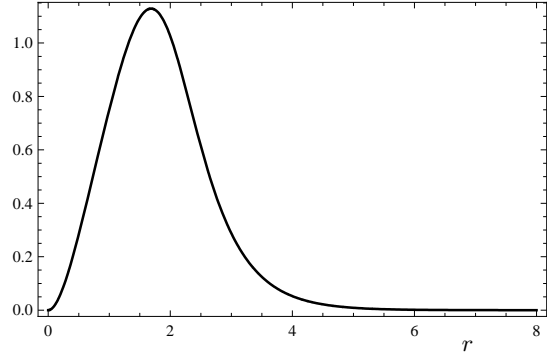}}
\put(185,6){$r$}
\end{picture} 
\caption{Profile of the integrand (in arbitrary units) appearing in Eq.~\eqref{eq:bdgammaA1} for $\Delta=3$ and $r_*=2$.}
\label{Fig:integrandprofile}
\end{center}
\end{figure}

The calculations of $\gamma_{V\,0}$ and $\gamma_{V\,1}$ are easily done, and yield
\beqs
\gamma_{V\,0}(r)&=&0\,,\\
b(r_1)\gamma_{V\,1}(r)&=&-r\,.
\eeqs
With this in place, using Eq.~\eqref{eq:bdgammaA1}, we can compute ($r_{\ast} \gg r_1 = 0$, $\Delta \gg 1$)
\beqs
b\delta\gamma_{A\,1} &\simeq& \,\frac{M_Z^2}{\varepsilon^2}\left(\frac{1}{2}+r_{\ast} +r_{\ast}^2 \right)e^{-2r_{\ast}}\,.
\eeqs
Fig.~\ref{Fig:integrandprofile} shows an example of the profile of the integrand that appears in this calculation. Hence, we find that
\beqs
b\gamma_{A\,1}&\simeq&-r+\frac{M_Z^2}{\varepsilon^2}\left(\frac{1}{2}+r_{\ast}+r_{\ast}^2\right)e^{-2 r_{\ast}}\,,
\eeqs
from which we have
\beqs
\hat{S} &\simeq& \cos^2\theta_W M_Z^2\left(\frac{1}{2}+r_{\ast}+r_{\ast}^2\right)e^{-2 r_{\ast}}\,.
\eeqs

Interestingly then, moving the symmetry breaking in the bulk results in 
a suppression of $\hat{S}$  controlled by $r_{\ast}$.
Even very moderate values of $r_{\ast}\simeq 1.5$ yield a suppression of $\hat{S}$
by a factor of $0.4$, which in turn means a lowering of the decay constant from $F\simeq 1.1$ TeV
to $F\simeq 0.7$ TeV (or a change from $a=v_W/F\simeq 0.22$ to $a=v_W/F\simeq 0.35$).
At the same time, the mass of the techni-rho mesons is lowered from $M_{\r}\gsim 2.4$ TeV
down to $M_{\r}\gsim 1.5 $ TeV.

\subsection{Summary}

In this section we reconsidered the model in~\cite{EPdilaton,LP}, but we changed the way in which electroweak symmetry breaking
is implemented, by assuming that the bulk scalar controlling the geometry is also 
the bulk Higgs field inducing electroweak symmetry breaking.
As expected on simple grounds, we find that because symmetry breaking is now due to effects that are localized
in the bulk near $r_{\ast}>0$, there is a parametric suppression of the $S$ parameter.
In turn, this allows to lower $\Lambda_0$
and hence the decay constant of the techni-dilaton $F$. This means that in models of this type the value
of $a=v_W/F$ is allowed to be larger than the $a=0.22$ which is found otherwise in most 
of the holographic models. But also that the
techni-rho mesons can be lighter than the generic models predict.

A big question remains open. 
Is this phenomenologically-motivated idea consistent with the gravity dual description of any 
strongly-coupled field theories? In order to answer to this question, one needs to 
discuss the truncation down to five dimensions of 10-dimensional backgrounds, 
retaining in the sigma-model spin-1 states and making the symmetries manifest.
One then has to ask whether there are cases where the $SU(2)_L\times U(1)_R$
symmetry of the Standard Model can be identified with a subgroup of the resulting five-dimensional gauge theory,
whether it is possible to trigger electroweak symmetry breaking via the VEV of some of the bulk scalars,
and whether it is possible to weakly gauge consistently  the dual global symmetry.
All of which constitutes a highly non-trivial program of explorations of possible 
string-theory models. However, the basic formal tools necessary are already available in the literature.

\section{Comparison to the experimental data}

In this section, we want to critically assess the present status of the experimental
measurements testing the couplings, the production cross-section and the  decay rates of the 
 boson with mass 125 GeV
discovered by the LHC experiments.
We do so by performing a simplified $\chi^2$ analysis of all the data.

\subsection{Definitions}

Let us start by summarizing the relevant physics process
for the Higgs searches at the LHC.
There are four main production mechanisms, gluon-gluon fusion (ggF),
vector-boson fusion (VBF) and associated production  with
a $W$ or a $Z$ boson (Vh) or with a top pair (tth). For $m_h\simeq 125$ GeV
we have the following  cross-sections~\cite{twiki7} at $7$ TeV:
\beqs
\sigma(pp\rightarrow h)_7^{SM}&=&15.3 \,{\rm pb}\, \pm 15\% \, \,,\\
\sigma(pp\rightarrow qqh)_7^{SM}&=&1.2 \,{\rm pb}\, \pm 2.7\%\,,\\
\sigma(pp\rightarrow Wh)_7^{SM}&=&0.57 \,{\rm pb}\, \pm 4.0\%\,,\\
\sigma(pp\rightarrow Zh)_7^{SM}&=&0.32 \,{\rm pb}\, \pm 5.0\%\,,\\
\sigma(pp\rightarrow tth)_7^{SM}&=&0.08 \,{\rm pb}\,\pm 17.0\%\,.
\eeqs
At $8$ TeV there is a small change~\cite{twiki8}, the cross-sections being larger:
\beqs
\sigma(pp\rightarrow h)_8^{SM}&=&19.5 \,{\rm pb}\, \pm 15\% \, \,,\\
\sigma(pp\rightarrow qqh)_8^{SM}&=&1.56 \,{\rm pb}\, \pm 2.9\%\,,\\
\sigma(pp\rightarrow Wh)_8^{SM}&=&0.70 \,{\rm pb}\, \pm 4.0\%\,,\\
\sigma(pp\rightarrow Zh)_8^{SM}&=&0.39 \,{\rm pb}\, \pm 5.0\%\,,\\
\sigma(pp\rightarrow tth)_8^{SM}&=&0.13 \,{\rm pb}\,\pm 17.0\%\,.
\eeqs

The production process is not directly observable at the LHC, in the sense that for any search strategy,
the events selected are going to come from an admixture of the various production mechanisms.
However, the ggF process has a cross-section that is so much larger than the others that
for searches involving untagged final states, we can safely assume that production is
entirely due to ggF.
For tagged processes, kinematical cuts allow to suppress the ggF contribution, but this is not going to be 
suppressed entirely. 

The number of events for  given processes of interest is proportional to the cross-sections for
the specific final states $R_i$, which are in general given by
\beqs
R_i&=&\sigma_i \,B_i\,,
\eeqs
with $\sigma_i$ the relevant production cross-section and $B_i$ the branching ratio.
For a standard-model Higgs with mass $m_h=125$ GeV the branching ratios are~\cite{twikiBR}:
\beqs
BR(h\rightarrow b\bar{b})_{SM}&=&58\%\,,\\
BR(h\rightarrow c\bar{c})_{SM}&=&2.7\%\,,\\
BR(h\rightarrow \tau^+{\tau}^-)_{SM}&=&6.4\%\,,\\
BR(h\rightarrow Z{Z}^{\ast})_{SM}&=&2.7\%\,,\\
BR(h\rightarrow W W^{\ast})_{SM}&=&21.6\%\,,\\
BR(h\rightarrow \gamma\gamma)_{SM}&=&0.22\%\,,\\
BR(h\rightarrow gg)_{SM}&=&8.5\%\,.
\eeqs

We {\it define} the simplest version of a  dilaton by the following properties. 
It is a parametrically light, scalar particle, with mass $m_d$ much lower than the mass 
of any other new particle besides the standard-model ones.
It couples to the standard-model particles 
via the stress-energy momentum tensor. Its leading-order couplings differ from those of the 
standard-model Higgs by the presence of three new parameters. 
\begin{itemize}
\item The decay constant $F\neq v_W$ of the dilaton means that the direct coupling to heavy vectors
($W$ and $Z$) and to standard-model fermions is rescaled by a parameter $a\equiv v_{W}/F$.
This coupling  modifies in a universal way all the production and decay processes,
and in sensible models $0<a<1$, with $a\rightarrow 1$ the limit in which the dilaton resembles closely the standard-model Higgs.
\item The coupling to photons of the standard-model Higgs is due to loops of $W$ and top quark.
For the dilaton, additional contributions come from new heavy particles. Because of the relative minus
sign in the contribution of vector bosons and fermions, a small-to-moderate number of new fermions would suppress this coupling, while a large number of new fermions leads to enhancement.
Hence, we parameterize this coupling by rescaling the SM one by a factor of $c_{\gamma}\,a$,
where $a$ has been introduced earlier, and $c_{\gamma}>0$ (notice that choosing negative values of $c_{\gamma}$ has the same effect, since this coupling always enters squared in physical processes).
\item The coupling to two gluons is very similar to that to photons, but for the fact that the new physics
can only enhance this coupling. We hence parameterize the new coupling by multiplying by a factor of
$c_{g}\,a$, with $c_g\geq 1$.
\end{itemize}

The parameter $c_{\gamma}$ does not affect the production cross-sections.
The parameter $a$ has a universal effect on all the cross-sections, while $c_g$
affects only the ggF process.
Hence
\beqs
\sigma(pp\rightarrow d)_{7,8}&=&a^2\,c_g^2\,\sigma(pp\rightarrow h)_{7,8}^{SM}\,,\\
\sigma(pp\rightarrow qqd)_{7,8}&=&a^2\,\sigma(pp\rightarrow qqh)_{7,8}^{SM}\,,\\
\sigma(pp\rightarrow Vd)_{7,8}&=&a^2\,\sigma(pp\rightarrow Vh)_{7,8}^{SM}\,,\\
\sigma(pp\rightarrow ttd)_{7,8}&=&a^2\,\sigma(pp\rightarrow tth)_{7,8}^{SM}\,.
\eeqs

The parameter $a$ does not affect any of the branching ratios, being universal.
However, $c_{g}$ and $c_{\gamma}$ do have an effect.
First of all, they rescale all the BR by a common factor of 
\beqs
\r&\equiv&\frac{1}{0.913+0.0022\,c_{\gamma}^2+0.085\,c_g^2}\,.
\eeqs
Notice that we are implicitly assuming that no new (invisible) decay channels exist,
aside from the SM ones.
Also, there is a direct effect on the decays to photons and gluons.
Hence, for a dilaton we have
\beqs
BR(d\rightarrow b\bar{b})&=&\r\,BR(h\rightarrow b\bar{b})_{SM}\,,\\
BR(d\rightarrow c\bar{c})&=&\r\,BR(h\rightarrow c\bar{c})_{SM}\,,\\
BR(d\rightarrow \tau^+{\tau}^-)&=&\r\,BR(h\rightarrow \tau^+\tau^-)_{SM}\,,\\
BR(d\rightarrow Z{Z}^{\ast})&=&\r\,BR(h\rightarrow ZZ^{\ast})_{SM}\,,\\
BR(d\rightarrow W W^{\ast})&=&\r\,BR(h\rightarrow WW^{\ast})_{SM}\,,\\
BR(d\rightarrow \gamma\gamma)&=&c_{\gamma}^2\,\r\,BR(h\rightarrow \gamma\gamma)_{SM}\,,\\
BR(d\rightarrow gg)&=&c_g^2\,\r\,BR(h\rightarrow gg)_{SM}\,.
\eeqs

Finally, the total width of the scalar is affected by $a$, $c_{\gamma}$ and $c_g$.
As we will see, $c_{\gamma}$ is always at most ${\cal O}(1)$, and hence does not really affect
the total width, due to the small branching ratio. However, parametrically large values of $c_g$
might result in a broadening of the width. Because the SM width is very small for $m_h\simeq 125$ GeV
($\Gamma_h\simeq$ few MeV) compared to
the energy resolution of the LHC experiments (in the GeV range), a large amount of freedom is left.
Nevertheless, we conservatively impose the bound $c_g<5/a$, which implies that we do not allow the width
of the dilaton to be more than a factor of $2-3$ larger than $\Gamma_h$.

We define the $\chi^2$ assuming that all the errors be Gaussian.
Notice also that some degree of imprecision has been introduced in the 
process of extracting the data from the literature, and that in doing so we also 
simplified the errors as symmetric.
Given the fact that at the moment all errors are quite large, that 
the distributions are unknown, but that 
there is a large number of independent measurements of the same quantities,
this is not going to affect the main results of the analysis.
We use the simple definition
\beqs
\chi^2 &=&\sum \frac{(r_{\rm th}-r_{\rm exp})^2}{\sigma^2}\,,
\eeqs
where $r$ refers to the ratio of the number of events to the SM prediction.
The function $\chi^2$ depends on $a$, $c_g$ and $c_{\gamma}$.
We hence use $\Delta \chi^2=3.53$ for $1\sigma$, $\Delta \chi^2=8.03$ for $2\sigma$
and $\Delta \chi^2 = 14.16$ for $3\sigma$~\cite{pdg}.

\subsection{Data analysis}

\begin{table}[t]
\begin{tabular}{||c|c|c|c||}
\hline\hline
&Process &  Comments & $({\rm ggF},{\rm VBF},{\rm Vh},{\rm tth})\%$ \cr
\hline\hline
CMS &&&\cr
\hline\hline
1&$h\rightarrow \gamma\gamma$ &  $7$ TeV Untagged $0$& $(61,17,19,3)$\cr
2&$h\rightarrow \gamma\gamma$ &  $7$ TeV Untagged $1$&$(88,6,6,1)$\cr
3&$h\rightarrow \gamma\gamma$ &  $7$ TeV Untagged $2$&$(91,4,4,0)$\cr
4&$h\rightarrow \gamma\gamma$ &  $7$ TeV Untagged $3$&$(91,4,4,0)$\cr
5&$h\rightarrow \gamma\gamma$ &  $7$ TeV Dijet Tag &$(27,73,1,0)$\cr
\hline
6&$h\rightarrow \gamma\gamma$ & $8$ TeV Untagged $0$& $(68,12,16,4)$\cr
7&$h\rightarrow \gamma\gamma$ &  $8$ TeV Untagged $1$& $(88,6,6,1)$\cr
8&$h\rightarrow \gamma\gamma$ &  $8$ TeV Untagged $2$& $(92,4,3,0)$\cr
9&$h\rightarrow \gamma\gamma$  &$8$ TeV Untagged $3$& $(92,4,4,0)$\cr
10&$h\rightarrow \gamma\gamma$ & $8$ TeV Dijet Tight& $(23,77,0,0)$\cr
11&$h\rightarrow \gamma\gamma$ &  $8$ TeV Dijet Loose $0$& $(53,45,2,0)$\cr
\hline
12&$h\rightarrow ZZ\rightarrow 4\ell$& $7$ TeV&\cr
13&$h\rightarrow ZZ\rightarrow 4\ell$& $8$ TeV&\cr
\hline
14&$h\rightarrow bb$ &  $7$ TeV Vh Tag &\cr
15&$h\rightarrow bb$ &  $8$ TeV Vh Tag &\cr
16&$h\rightarrow bb$ &  $8$ TeV tth Tag &\cr
\hline
17&$h\rightarrow \tau\tau$ & $7$ TeV $0/1$ jet &\cr
18&$h\rightarrow \tau\tau$ &  $8$ TeV $0/1$ jet &\cr
19&$h\rightarrow \tau\tau$ &  $7$ TeV VBF Tag &\cr
20&$h\rightarrow \tau\tau$ & $8$ TeV VBF Tag &\cr
21&$h\rightarrow \tau\tau$ &  $7$ TeV Vh Tag &\cr
\hline
22&$h\rightarrow WW$ &  $7$ TeV $0/1$ jet &\cr
23&$h\rightarrow WW$ &  $8$ TeV $0/1$ jet &\cr
24&$h\rightarrow WW$ & $7$ TeV VBF Tag &\cr
25&$h\rightarrow WW$ &  $8$ TeV VBF Tag &\cr
26&$h\rightarrow WW$ &  $7$ TeV Vh Tag &\cr
\hline\hline
ATLAS &&&\cr
\hline\hline
27&$h\rightarrow \gamma\gamma$ &  $7$ TeV & \cr
28&$h\rightarrow \gamma\gamma$ &  $8$ TeV &\cr
\hline
29&$h\rightarrow ZZ\rightarrow 4\ell$&$7$ TeV&\cr
30&$h\rightarrow ZZ\rightarrow 4\ell$&$8$ TeV&\cr
\hline
31&$h\rightarrow bb$ & $7$ TeV &\cr
\hline
32&$h\rightarrow \tau\tau$ & $7$ TeV  &\cr
\hline
33&$h\rightarrow WW$ &  $7$ TeV &\cr
34&$h\rightarrow WW$ &  $8$ TeV &\cr
\hline\hline
TeVatron &&&\cr
\hline\hline
35 & $h\rightarrow \gamma\gamma$  &  & \cr
36 & $h\rightarrow bb$  && \cr
37 & $h\rightarrow WW$ && \cr
\hline\hline
\end{tabular}
\caption{Categories of data used in the analysis.}
\label{Fig:data}
\end{table}

\begin{figure}[t]
\begin{center}
\begin{picture}(500,160)
\put(115,10){\includegraphics[height=5.5cm]{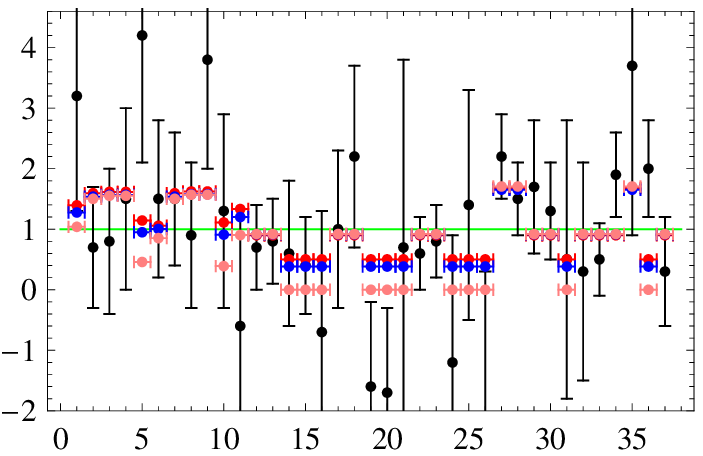}}
\put(110,140){$r$}
\end{picture} 
\caption{Rates $r$ as measured by CMS, ATLAS and TeVatron (black), normalized to the SM  prediction,
and compared to the SM (green), to the best fit from the dilaton of all the data (red),
 of all the data excluding the 7 TeV $\gamma\gamma$ dijet tagged, point 5  (blue), and of all the data excluding both this  and the bb TeVatron measurement, point 36 (pink).}
\label{Fig:TOTrates}
\end{center}
\end{figure}

The measurements we include in our analysis 
are summarized in Table~\ref{Fig:data} and in Fig.~\ref{Fig:TOTrates}.
When known, we explicitly write the composition of the data. 
Otherwise, we assume $100\%$ purity in the samples
for ggF, VBF, Vh and tth processes.
The CMS results are taken from~\cite{CMS,newCMS}.
The ATLAS results are from~\cite{EGMT,newATLAS}.
The TeVatron results are taken from~\cite{EGMT}.

\begin{figure}[t]
\begin{center}
\begin{picture}(600,140)
\put(10,10){\includegraphics[height=4.5cm]{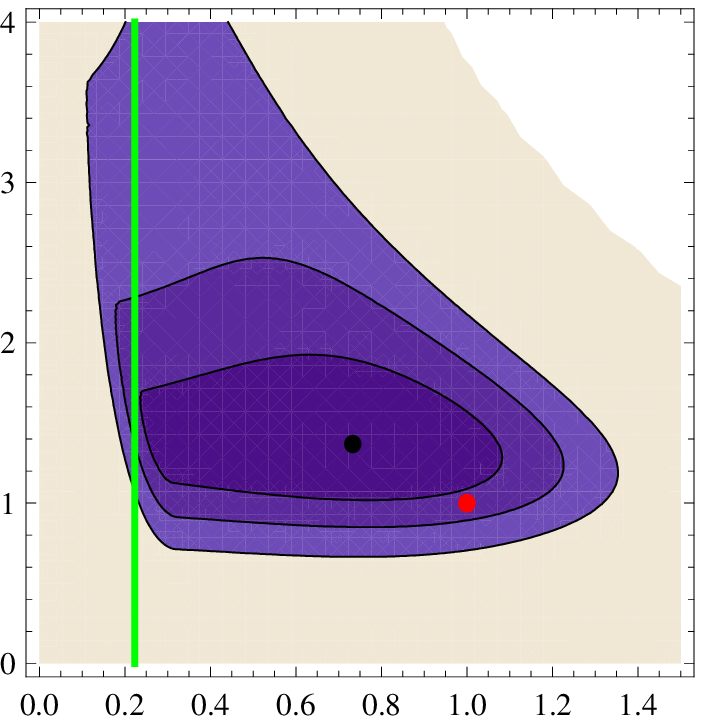}}
\put(190,10){\includegraphics[height=4.5cm]{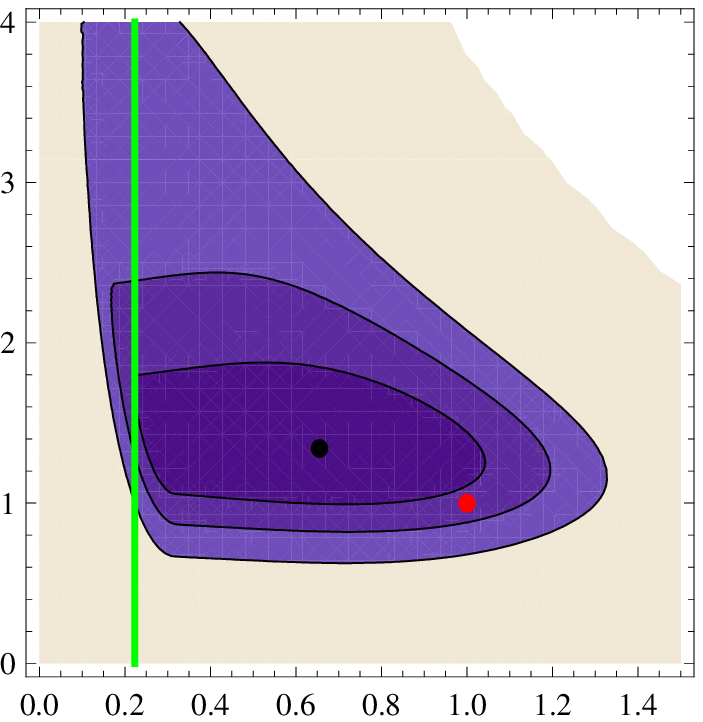}}
\put(370,10){\includegraphics[height=4.5cm]{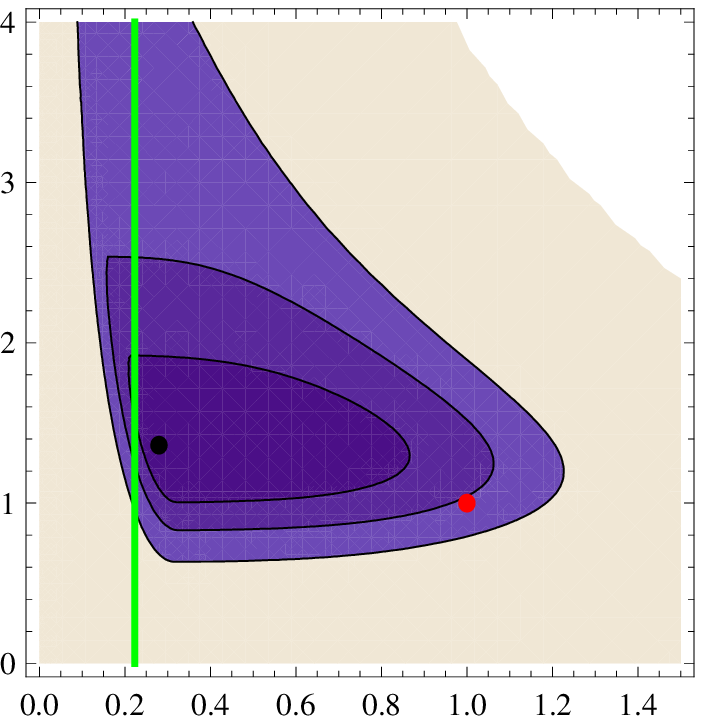}}
\put(0,120){$c_{\gamma}$}
\put(180,120){$c_{\gamma}$}
\put(360,120){$c_{\gamma}$}
\put(120,6){$a$}
\put(300,6){$a$}
\put(480,6){$a$}
\end{picture} 
\caption{Numerical result of $\chi^2$ analysis of CMS+ATLAS+TeVatron data. 
Shown is the best fit point (black dot), the SM prediction (red dot), and the
 $1\sigma$, $2\sigma$ and $3\sigma$ contours obtained by minimizing in respect to $c_g$.
The left panel uses all the data, in the central we omitted the $h\rightarrow \gamma\gamma$
dijet tagged events at 7 TeV from CMS, 
and in the right panel we omitted both this and the bb data from TeVatron. 
The lines correspond to $a\simeq 0.22$}
\label{Fig:TOT}
\end{center}
\end{figure}

We show the result of the 3-parameter fit  in Fig.~\ref{Fig:TOT}.
In order to provide a better illustration of the results, we plot in Fig.~\ref{Fig:TOTrates} 
the data against the SM prediction and
against the three best fits shown in the three panels of Fig.~\ref{Fig:TOT}.

Before giving details about the analysis, a comment for the reader to guide reading the plots
in Fig.~\ref{Fig:TOT}.
The sharp edge on the contours for small $a$ comes from the fact that for very small $a$ 
the $\chi^2$
has a runaway behavior in the $c_g$ direction, and effectively 
means that it is not possible to fit the data with arbitrarily small choices of $a$. 
The fact that the holographic techni-dilaton prediction $a\simeq 0.22$ lies so close to the point were 
this runaway behavior appears is a  reason for moderate concern within this type of models.

Let us now comment on the results of the $\chi^2$ analysis, displayed in Fig.~\ref{Fig:TOT}. 
The minimum $\chi^2$ of all data is $\chi^2=24.2$ for $(a,c_{\gamma},c_g)=(0.73,1.37,1.34)$,
while the SM $(a,c_{\gamma},c_g)=(1,1,1)$  yields $\chi^2=30.0$, within $2\sigma$. 
This is shown in the left panel of Fig.~\ref{Fig:TOT}. The 2-dimensional 
figure has been obtained by minimizing, for
every choice of $a$ and $c_{\gamma}$, the $\chi^2$ as a function of $c_g$ 
(and with the constraint $c_g<5/a$). If we set $a=1=c_{\gamma}$,
the minimum of the $\chi^2$ is at $c_g\simeq 1.08$, which justifies the fact that
we show the SM prediction on this plot.
The holographic techni-dilaton with $a=0.22$
barely touches the $1\sigma$ contour, for $c_{\gamma}\simeq 1.6$ and large $c_g$.

The reason why some moderate amount of  tension with the holographic  techni-dilaton
hypothesis is present
 can be seen in Fig.~\ref{Fig:TOTrates}: points 5 and 36 are substantially higher
than the SM prediction, while the holographic techni-dilaton would suppress these two processes
in respect to the SM Higgs boson case.
The former is the $7$ TeV Dijet Tag $\gamma\gamma$ category from CMS.
Here, there might be a larger contamination from ggF than what reported by CMS.
While this statement cannot be tested with present knowledge, notice that 
such large excess is absent in the $8$ TeV data from CMS itself (points 10 and 11).
Category 36 represents the bb searches at TeVatron. Here the production 
process is Vh, which would be  suppressed for a dilaton.
TeVatron reports a very large excess even in respect to the SM Higgs hypothesis.
However, neither CMS nor ATLAS see such an effect at present (points 14,15,16 and 31).
In order to assess how significant these two high measurements
are in the context of the global analysis, we perform the exercise of excluding them from the analysis itself.

 By excluding only the dijet tagged $\gamma\gamma$ at $7$ TeV from CMS, we obtain a somewhat
 lower  $\chi^2=21.9$
 for $(a,c_{\gamma},c_g)=(0.65,1.34,1.54)$. In this case the SM  improves to $\chi^2=27.7$.
 This is a  marginal change, with the $a=0.22$ choice allowed inside the $1\sigma$ contour
 for a range of $c_{\gamma}=1.4-1.8$.
 As a third and last exercise, excluding both the dijet tagged photons at $7$ TeV from CMS and the TeVatron 
 bb lowers substantially  the $\chi^2$  to
 $\chi^2=16.7$ for $(a,c_{\gamma})=(0.28,1.36)$. In this case though, there is no preferred value for 
 $c_g$, which has to be large for the best-fit choices of $a$ and $c_{\gamma}$ 
 and can be taken to be at the upper bound we applied ($c_g=5/a$). 
 In this case the SM has $\chi^2=26.1$, and lies just outside the $2\sigma$ region.
 Notice from the figure how close the best fit is to the holographic techni-dilaton with $a=0.22$.

A few more comments about Fig.~\ref{Fig:TOTrates}. Notice how the three fits agree for all the ggF
processes. Some tension appears for data for which substantial components of VBF, Vh or tth
are assumed. In particular, notice how the final  fit works acceptably well for almost all data, 
with the fit being either within $1\sigma$ of the experimental error, or very close to it.
The only exceptions 
are the data points 5 and 36 that we excluded in some of the fits.
Notice however that even including these data points in the analysis {\it does not}
bring the fit anywhere near the $1\sigma$ error bars for points 5 and 36. 
If such results are going to be confirmed with more statistics (and smaller error bars), the holographic techni-dilaton hypothesis 
would be strongly disfavored, but for the time being the status is inconclusive.
Finally, notice that the only other points which are visibly affected by the change of fitting procedure 
(i.~e. by excluding points 5 and 36) are points 
19 and 20, the VBF tagged $\tau\tau$ events from CMS. 
These two points are clearly anomalous, the rate to the SM Higgs ratio being negative
(indicating a negative fluctuation in the background). 
However the selective fit with points 5 and 36 excluded gets quite close to the $1\sigma$ region
for these two processes. But again, the uncertainties are so large that this is
inconclusive, at present.

\subsection{Summary}

We conclude this phenomenological section by summarizing what we did and what we learned.
We fitted CMS,  ATLAS and TeVatron measurements of the event rates
for various production and decay processes  to a model in which 
a scalar boson of mass $125$ GeV has couplings to the other SM particles controlled by
three free parameters $0<a\leq 1$, $c_{\gamma}\geq 0$ and $c_g\geq 1$.
The SM Higgs is recovered for $a=1=c_g=c_{\gamma}$, while the holographic technidilaton
prefers  $a\lsim 0.22$, with $c_g$ and $c_{\gamma}$ free parameters.

We find that the best fit of all the data is given by  $(a,c_{\gamma},c_g)=(0.73,1.37,1.34)$.
The fact that $a<1$ results in a suppression of all the rates in respect to the SM.
This is compensated in the case of ggF production by taking $c_g$ large.
And finally by choosing $c_{\gamma}\simeq 1.37$ the fit agrees well with the data 
on $\gamma\gamma$ from ggF.
The SM is just inside the $2\sigma$ region identified by this $\chi^2$,
while  the holographic techni-dilaton with $a=0.22$  touches the $1\sigma$ region. 
Hence there is no clear indication in the data that either of the two hypotheses is favored nor disfavored.

However, a tension is present within the data itself, with most of the measurements 
affected by VBF, Vh and tth production showing some suppression in respect to the SM,
with the noticeable exceptions of the dijet tagged $\gamma\gamma$ 
events collected by CMS at $7$ TeV,
and the bb events from TeVatron. Given the large uncertainties, no firm conclusion can be drawn at the moment.
Yet, it is interesting to notice that by excluding these two 
data points, but using all the other 35, yields a $\chi^2$ which improves from  
$\chi^2=24.2$ to $\chi^2=16.7$. In doing so, the best fit now has $a=0.28$ and $c_{\gamma}=1.36$,
while $c_g$ is large, a set of choices that is very close to the expectations from holographic techni-dilaton.

The conclusion is that the holographic techni-dilaton  fits the data (marginally) better than the SM, but the present data is not able to distinguish unambiguously the two:
\begin{itemize}
\item the holographic techni-dilaton explains without difficulty the enhanced $\gamma\gamma$ signal that all
the experimental collaborations reported,
\item the holographic techni-dilaton would prefer a strong suppression of VBF, Vh and tth 
production mechanisms, which agrees with some but not all the experimental measurements,
 in particular the TeVatron Vbb and CMS 7 TeV tagged dijets $\gamma\gamma$ category 
 have too high rates, unfortunately not clearly confirmed by other measurements, 
\item if the contamination from ggF production to  Vbb and to  $jj\gamma\gamma$ processes
is somewhat larger than what we assumed ($0\%$ for the former, $27\%$ for the latter) 
the tension in the data would soften,
\item the ggF production is at the same time suppressed by $a<1$ and enhanced by $c_g>1$,
in such a way that the overall number of Higgs-like events from ggF is close to the SM prediction 
(it is a completely open question whether this is implemented in a sensible way in a specific UV-complete model, or whether this is the result of a
bizarre conspiracy between independent free parameters, but see~\cite{MY}),
\item the special choice $a=0.22$ is always marginally allowed to be in the $1\sigma$ region, for all the
different choices we made, but also at the price of having to use very large values for $c_g$,
\item a modest modification of the properties of the holographic techni-dilaton and their correlation 
with precision physics (along the lines described in the
previous sections of this paper) would allow to
fit very well the data, provided in this way one could tolerate values of $a\gsim 0.3$.
\end{itemize}

With more data, it should be possible to clarify these small tensions, 
which are all at the $1 \sigma$ or $2 \sigma$ level at most. 
In particular, we need to watch in future data released by the LHC experiments
whether the excess in $\gamma\gamma$ persists (favoring the dilaton),
but also whether some unambiguous evidence for production 
mechanisms other that gluon-gluon fusion  emerges, which would require a low
value for the dilaton decay constant,  and hence disfavor models with small $a$.

\section{Conclusions}

We discussed the physics of the holographic techni-dilaton,
a composite scalar state emerging from a strongly-coupled field-theory
mechanism of electroweak symmetry breaking the 
properties of which are computed using the ideas of gauge-gravity dualities,
in terms of a five-dimensional sigma-model coupled to gravity.

We started from the well-known fact that such techni-dilaton
has a mass which depends critically on the (model-dependent) details of the strongly-coupled sector,
which  can be compatible with the 125 GeV mass of the new boson.
Also, it is well-known that because of new physics contributions in the loops,
the holographic techni-dilaton has (model-dependent) 
enhanced couplings to two photons and to two gluons in
respect to the Higgs boson of the minimal version of the Standard Model.
We also recall the following three points.
\begin{itemize}
\item Computing the decay constant of the holographic techni-dilaton 
yields a remarkably universal result, with $F/\Lambda_0\simeq \sqrt{3/2}$,
in terms of the overall scale $\Lambda_0$ of the strong dynamics.
\item If one models electroweak symmetry by allowing the electroweak gauge bosons to
propagate in the bulk of the fifth dimension, and models electroweak symmetry breaking in terms of IR-localized physics,
the bounds from electroweak precision measurements yield a remarkably universal 
bound $\Lambda_0\gsim 900$ GeV, which translates into an equally model-independent bound
on the mass of the lightest techni-rho meson $M_{\r}\gsim 2.4$ TeV.
In turns, these two points translate into the fact that $F\gsim 1.1$ TeV, and hence all the couplings
of the techni-dilaton are suppressed by a factor of $a=v_{W}/F\lsim 0.22$ in respect to the Higgs boson.
\item The experimental results from LHC and TeVatron show an enhanced rate into two photons, which
favors the holographic techni-dilaton hypothesis, together with hints of significant components 
of the Higgs-like signal coming from production mechanisms other than gluon-gluon fusion,
a fact that disfavors the techni-dilaton hypothesis.
Once put together in a global fit, the two facts make the techni-dilaton just marginally favored over the 
Higgs boson hypothesis, but the status is presently inconclusive.
\end{itemize}

In this paper we started to question the validity of these three points and asked how to improve our general
understanding of the physics they descend from.
We built new models in which the geometry is significantly different from AdS, and confirmed 
that even in these models $F/\Lambda_0\simeq 1.2$.
Whether counterexamples can be found is still an open problem, but it would require
to analyze models where the departures from AdS are very large, and effectively dominate the dynamics.

We considered the case in which electroweak symmetry breaking emerges from bulk effects, 
as opposed to being localized in the deep-IR. We find that there is a parametric suppression of the $S$ parameter in this
case, and hence the scale $\Lambda_0$ can be lowered,  in turn  allowing for larger values of $a$
which would soften the suppression of VBF, Vh and tth production mechanisms.
Whether such a scenario can be realized in a consistent way within string-theory constructions is 
a completely open problem.

We also performed a global fit of all the data on Higgs searches relevant to the 125 GeV mass range.
We showed that while the techni-dilaton is marginally favored by the data over the SM Higgs boson,
there exists some tension in the data. We also showed that for the most part this tension can be removed
if one excludes from the analysis the TeVatron bb search and the 7 TeV CMS $jj\gamma\gamma$ search.
We  do not know whether this is legitimate or not, but it is clear that a more robust confirmation of these two
results from other collaborations would be needed in order to clearly assess whether the tension
with the techni-dilaton hypothesis is real or just an unfortunate result of the present, large uncertainties in the measurements.
In particular, it will be very important to see what the $jj\gamma\gamma$, $Vbb$, $jjWW$ and $jj\tau\tau$ 
analysis of  future CMS and ATLAS data show.

In conclusion, the holographic techni-dilaton is a very plausible candidate for interpreting 
the recently discovered 125 GeV mass boson, competitive  with the Higgs boson of the minimal version of the Standard Model.
Both from a theoretical and from an experimental viewpoint, much more work is needed in order to disentangle the two.
But while the road to go might be long and difficult, the directions to follow  are  quite clear.

\vspace{1.0cm}
\begin{acknowledgments}
The work of MP is supported in part by WIMCS and by the STFC grant ST/J000043/1.

\end{acknowledgments}



\begin{thebibliography}{99}

\bibitem{ATLASJuly2012}
https://indico.cern.ch/getFile.py/access?contribId=1\&resId=1\&materialId=slides\&confId=197461

\bibitem{CMSJuly2012}
https://cms-docdb.cern.ch/cgi-bin/PublicDocDB/RetrieveFile?docid=6125\&filename=CMS\_4July2012\_Incandela.pdf


\bibitem{LLS}
 I.~Low, J.~Lykken and G.~Shaughnessy,
  arXiv:1207.1093 [hep-ph].
  
\bibitem{Corbett:2012dm} 
  T.~Corbett, O.~J.~P.~Eboli, J.~Gonzalez-Fraile and M.~C.~Gonzalez-Garcia,
  arXiv:1207.1344 [hep-ph].
  
  \bibitem{GKRS}
  P.~P.~Giardino, K.~Kannike, M.~Raidal and A.~Strumia,
  arXiv:1207.1347 [hep-ph].
  \bibitem{EY}
   J.~Ellis and T.~You,
  arXiv:1207.1693 [hep-ph].
  
\bibitem{Montull:2012ik} 
  M.~Montull and F.~Riva,
  arXiv:1207.1716 [hep-ph].

\bibitem{EGMT}
 J.~R.~Espinosa, C.~Grojean, M.~Muhlleitner and M.~Trott,
  arXiv:1207.1717 [hep-ph].

\bibitem{CFKVZ}
   D.~Carmi, A.~Falkowski, E.~Kuflik, T.~Volansky and J.~Zupan,
  arXiv:1207.1718 [hep-ph];

\bibitem{Bertolini:2012gu} 
  D.~Bertolini and M.~McCullough,
  arXiv:1207.4209 [hep-ph].

  \bibitem{MY}
 S.~Matsuzaki and K.~Yamawaki,
  arXiv:1207.5911 [hep-ph].

\bibitem{TC}
 S.~Weinberg,
  Phys.\ Rev.\ D {\bf 19}, 1277 (1979);
L.~Susskind,
  Phys.\ Rev.\ D {\bf 20}, 2619 (1979);
 S.~Weinberg,
  Phys.\ Rev.\ D {\bf 13}, 974 (1976).


\bibitem{reviews}
 R.~S.~Chivukula,
  arXiv:hep-ph/0011264;
K.~Lane,
  arXiv:hep-ph/0202255,
 C.~T.~Hill and E.~H.~Simmons,
  Phys.\ Rept.\  {\bf 381}, 235 (2003)
  [Erratum-ibid.\  {\bf 390}, 553 (2004)]
  [arXiv:hep-ph/0203079];
  A.~Martin,
  arXiv:0812.1841 [hep-ph];
   F.~Sannino,
  arXiv:0911.0931 [hep-ph];
   M.~Piai,
  Adv.\ High Energy Phys.\  {\bf 2010}, 464302 (2010)
  [arXiv:1004.0176 [hep-ph]].

 \bibitem{Peskin}
 M.~E.~Peskin and T.~Takeuchi,
  Phys.\ Rev.\ D {\bf 46}, 381 (1992).
  
  

\bibitem{Barbieri}
R.~Barbieri, A.~Pomarol, R.~Rattazzi and A.~Strumia,
  Nucl.\ Phys.\ B {\bf 703}, 127 (2004)
  [arXiv:hep-ph/0405040].


\bibitem{WTC}
 B.~Holdom,
  Phys.\ Lett.\ B {\bf 150}, 301 (1985);
    K.~Yamawaki {\it et al.}
  Phys.\ Rev.\ Lett.\  {\bf 56}, 1335 (1986);
T.~W.~Appelquist {\it et al.}
  Phys.\ Rev.\ Lett.\  {\bf 57}, 957 (1986).





    \bibitem{dilaton1}
    M.~Bando {\it et al.}
  Phys.\ Lett.\  B {\bf 178}, 308 (1986);
  Phys.\ Rev.\ Lett.\  {\bf 56}, 1335 (1986);
    \bibitem{dilaton2}
  W.~A.~Bardeen {\it et al.}
  Phys.\ Rev.\ Lett.\  {\bf 56}, 1230 (1986);
    \bibitem{dilaton3}
    B.~Holdom and J.~Terning,
  Phys.\ Lett.\  B {\bf 187}, 357 (1987);
  Phys.\ Lett.\  B {\bf 200}, 338 (1988).
  
\bibitem{dilatonpheno}
W.~D.~Goldberger {\it et al.}
  Phys.\ Rev.\ Lett.\  {\bf 100}, 111802 (2008);
  and 
  L.~Vecchi,
  arXiv:1002.1721 [hep-ph].

\bibitem{dilaton4}
D.~D.~Dietrich, F.~Sannino and K.~Tuominen,
  Phys.\ Rev.\  D {\bf 72}, 055001 (2005)
  [arXiv:hep-ph/0505059].
 T.~Appelquist and Y.~Bai,
  arXiv:1006.4375 [hep-ph].
  L.~Vecchi,
  arXiv:1007.4573 [hep-ph];
   K.~Haba, S.~Matsuzaki and K.~Yamawaki,
  Phys.\ Rev.\  D {\bf 82}, 055007 (2010)
  [arXiv:1006.2526 [hep-ph]];
        M.~Hashimoto and K.~Yamawaki,
  arXiv:1009.5482 [hep-ph].

\bibitem{dilaton5D}
 W.~D.~Goldberger and M.~B.~Wise,
  Phys.\ Lett.\  B {\bf 475}, 275 (2000)
  [arXiv:hep-ph/9911457];
  O.~DeWolfe, D.~Z.~Freedman, S.~S.~Gubser and A.~Karch,
  Phys.\ Rev.\  D {\bf 62}, 046008 (2000)
  [arXiv:hep-th/9909134].
C.~Csaki, M.~L.~Graesser and G.~D.~Kribs,
  Phys.\ Rev.\  D {\bf 63}, 065002 (2001)
  [arXiv:hep-th/0008151];
  L.~Kofman, J.~Martin and M.~Peloso,
  Phys.\ Rev.\  D {\bf 70}, 085015 (2004)
  [arXiv:hep-ph/0401189].
  
  \bibitem{dilatonnew}
   K.~Cheung and T.~-C.~Yuan,
  Phys.\ Rev.\ Lett.\  {\bf 108}, 141602 (2012)
  [arXiv:1112.4146 [hep-ph]];
  S.~Matsuzaki and K.~Yamawaki,
  arXiv:1206.6703 [hep-ph].
  See also C.~D.~Carone,  
arXiv:1206.4324 [hep-ph].

\bibitem{EPdilaton}
D.~Elander and M.~Piai
  arXiv:1112.2915 [hep-ph].

\bibitem{LP}
 R.~Lawrance and M.~Piai,
  arXiv:1207.0427 [hep-ph].

\bibitem{ENP}
 D.~Elander, C.~Nunez and M.~Piai,
  Phys.\ Lett.\ B {\bf 686}, 64 (2010)
  [arXiv:0908.2808 [hep-th]].




\bibitem{AdSCFT}
  J.~M.~Maldacena,
  Adv.\ Theor.\ Math.\ Phys.\  {\bf 2}, 231 (1998)
  [Int.\ J.\ Theor.\ Phys.\  {\bf 38}, 1113 (1999)]
  [arXiv:hep-th/9711200];
  S.~S.~Gubser, I.~R.~Klebanov and A.~M.~Polyakov,
  Phys.\ Lett.\  B {\bf 428}, 105 (1998)
  [arXiv:hep-th/9802109];
  E.~Witten,
  Adv.\ Theor.\ Math.\ Phys.\  {\bf 2}, 253 (1998)
  [arXiv:hep-th/9802150].


\bibitem{reviewAdSCFT}
 O.~Aharony, S.~S.~Gubser, J.~M.~Maldacena, H.~Ooguri and Y.~Oz,
  Phys.\ Rept.\  {\bf 323}, 183 (2000)
  [arXiv:hep-th/9905111].





\bibitem{AdSTC}
   D.~K.~Hong and H.~U.~Yee,
  Phys.\ Rev.\  D {\bf 74}, 015011 (2006)
  [arXiv:hep-ph/0602177];
 M.~Piai,
  arXiv:hep-ph/0608241,
  arXiv:hep-ph/0609104,
  arXiv:0704.2205 [hep-ph];
    K.~Haba, S.~Matsuzaki and K.~Yamawaki,
  arXiv:0804.3668 [hep-ph];
M.~Round,
  arXiv:1003.2933 [hep-ph].
J.~Hirn and V.~Sanz,
  Phys.\ Rev.\ Lett.\  {\bf 97}, 121803 (2006)
  [arXiv:hep-ph/0606086],
  JHEP {\bf 0703}, 100 (2007)
  [arXiv:hep-ph/0612239];
     C.~D.~Carone, J.~Erlich and J.~A.~Tan,
  arXiv:hep-ph/0612242;
M~Fabbrichesi, M.~Piai, L.~Vecchi
arXiv:0804.0124 [hep-ph];
  J.~Hirn, A.~Martin and V.~Sanz,
  arXiv:0807.2465 [hep-ph];
   M.~Round,
  Phys.\ Rev.\ D\ {\bf 84}, 013012  (2011)
  [arXiv:1104.4037 [hep-ph]].



\bibitem{stringS}
 C.~D.~Carone, J.~Erlich and M.~Sher,
  Phys.\ Rev.\ D {\bf 76}, 015015 (2007)
  [arXiv:0704.3084 [hep-th]];
   T.~Hirayama and K.~Yoshioka,
  JHEP {\bf 0710}, 002 (2007)
  [arXiv:0705.3533 [hep-ph]].
   O.~Mintakevich and J.~Sonnenschein,
  JHEP {\bf 0907}, 032 (2009)
  [arXiv:0905.3284 [hep-th]];
    L.~Anguelova,
  Nucl.\ Phys.\ B\ {\bf 843}, 429  (2011)
  [arXiv:1006.3570 [hep-th]];
   L.~Anguelova, P.~Suranyi and L.~C.~R.~Wijewardhana,
  Nucl.\ Phys.\ B\ {\bf 852}, 39  (2011)
  [arXiv:1105.4185 [hep-th]];
  

\bibitem{stringWTC}
C.~Nunez, I.~Papadimitriou and M.~Piai,
  arXiv:0812.3655 [hep-th];
  D.~Elander, C.~Nunez and M.~Piai,
  Phys.\ Lett.\  B {\bf 686}, 64 (2010)
  [arXiv:0908.2808 [hep-th]];
   C.~Nunez, M.~Piai and A.~Rago,
  arXiv:0909.0748 [hep-th].
D.~Elander, J.~Gaillard, C.~Nunez and M.~Piai,
  JHEP\ {\bf 1107}, 056  (2011)
  [arXiv:1104.3963 [hep-th]].
 L.~Anguelova, P.~Suranyi and L.~C.~R.~Wijewardhana,
  Nucl.\ Phys.\ B {\bf 862}, 671 (2012)
  [arXiv:1203.1968 [hep-th]].

\bibitem{stringWTC2}
 S.~P.~Kumar, D.~Mateos, A.~Paredes and M.~Piai,
  JHEP {\bf 1105}, 008 (2011)
  [arXiv:1012.4678 [hep-th]];
D.~Kutasov, J.~Lin and A.~Parnachev,
  Nucl.\ Phys.\ B {\bf 863}, 361 (2012)
  [arXiv:1201.4123 [hep-th]].

\bibitem{Rattazzi-Zaffaroni} 
  R.~Rattazzi and A.~Zaffaroni,
  JHEP {\bf 0104}, 021 (2001)
  [hep-th/0012248].

\bibitem{FPV}
 M.~Fabbrichesi, M.~Piai and L.~Vecchi,
  Phys.\ Rev.\ D {\bf 78}, 045009 (2008)
  [arXiv:0804.0124 [hep-ph]].

\bibitem{GPPZ}
 L.~Girardello, M.~Petrini, M.~Porrati and A.~Zaffaroni,
  JHEP {\bf 9812}, 022 (1998)
  [hep-th/9810126];
  J.~Distler and F.~Zamora,
  Adv.\ Theor.\ Math.\ Phys.\  {\bf 2}, 1405 (1999)
  [hep-th/9810206].
 L.~Girardello, M.~Petrini, M.~Porrati and A.~Zaffaroni,
  Nucl.\ Phys.\  B {\bf 569}, 451 (2000)
  [arXiv:hep-th/9909047].

\bibitem{GPPZspectrum}
  R.~Apreda, D.~E.~Crooks, N.~J.~Evans and M.~Petrini,
  JHEP {\bf 0405}, 065 (2004)
  [hep-th/0308006].
 M.~Bianchi, M.~Prisco and W.~Mueck,
  JHEP {\bf 0311}, 052 (2003)
  [hep-th/0310129].
 W.~Mueck and M.~Prisco,
  JHEP {\bf 0404}, 037 (2004)
  [hep-th/0402068].

\bibitem{hyperscaling}
 X.~Dong, S.~Harrison, S.~Kachru, G.~Torroba and H.~Wang,
  arXiv:1201.1905 [hep-th].

\bibitem{SS}
  E.~Witten,
  Adv.\ Theor.\ Math.\ Phys.\  {\bf 2}, 505 (1998)
  [arXiv:hep-th/9803131],
  T.~Sakai and S.~Sugimoto,
  Prog.\ Theor.\ Phys.\  {\bf 113}, 843 (2005)
  [arXiv:hep-th/0412141].


  \bibitem{HR}
K.~Skenderis,
  Class.\ Quant.\ Grav.\  {\bf 19}, 5849 (2002)
  [arXiv:hep-th/0209067];
  I.~Papadimitriou and K.~Skenderis,
  arXiv:hep-th/0404176.

\bibitem{BHM}
 M.~Bianchi, M.~Prisco and W.~Mueck,
  JHEP {\bf 0311}, 052 (2003)
  [arXiv:hep-th/0310129];
  M.~Berg, M.~Haack and W.~Mueck,
  Nucl.\ Phys.\  B {\bf 736}, 82 (2006)
  [arXiv:hep-th/0507285];
    M.~Berg, M.~Haack and W.~Mueck,
  Nucl.\ Phys.\  B {\bf 789}, 1 (2008)
  [arXiv:hep-th/0612224].

    
\bibitem{EP}
 D.~Elander,
  JHEP {\bf 1003}, 114 (2010)
  [arXiv:0912.1600 [hep-th]];
  D.~Elander and M.~Piai,
  JHEP {\bf 1101}, 026 (2011)
  [arXiv:1010.1964 [hep-th]].


\bibitem{PW}
 K.~Pilch and N.~P.~Warner,
  Adv.\ Theor.\ Math.\ Phys.\  {\bf 4}, 627 (2002)
  [hep-th/0006066].

\bibitem{PS}
 J.~Polchinski and M.~J.~Strassler,
  hep-th/0003136.



\bibitem{twiki7}
https://twiki.cern.ch/twiki/bin/view/LHCPhysics/CERNYellowReportPageAt7TeV

\bibitem{twiki8}
https://twiki.cern.ch/twiki/bin/view/LHCPhysics/CERNYellowReportPageAt8TeV

\bibitem{twikiBR}
https://twiki.cern.ch/twiki/bin/view/LHCPhysics/CERNYellowReportPageBR

\bibitem{CMS}
http://cms.web.cern.ch/org/cms-higgs-results

\bibitem{pdg}
 J. Beringer et al. (Particle Data Group), Phys. Rev. D86, 010001 (2012)

\bibitem{newATLAS} 
  G.~Aad {\it et al.}  [The ATLAS Collaboration],
  arXiv:1207.7214 [hep-ex].

\bibitem{newCMS} 
  S.~Chatrchyan {\it et al.}  [The CMS Collaboration],
  arXiv:1207.7235 [hep-ex].

\end{thebibliography}
\end{document}